\documentclass[
  journal=largetwo,
  manuscript=article-type,
  year=2020,
  volume=37,
]{cup-journal}

\usepackage{amsmath}
\usepackage[nopatch]{microtype}
\usepackage{booktabs}
\usepackage{hyperref}
\usepackage{xspace}

\usepackage{amssymb}	
\usepackage{multicol}        
\usepackage{bm}		
\usepackage{pdflscape}	
\usepackage{multirow}
\usepackage{arydshln}
\usepackage{float}
\usepackage{subfigure}

\newcommand{\kms}{\,km\,s$^{-1}$} 
\usepackage{array}
\newcolumntype{H}{>{\setbox0=\hbox\bgroup}c<{\egroup}@{}}

\title{The Dusty Aftermath of a Rapid Nova: V5579 Sgr}

\author{A. Raj}
\affiliation{Indian Centre for Space Physics, 466 Barakhola, Netai Nagar, Kolkata, 700099, West Bengal, India}
\email[A. Raj]{ashishpink@gmail.com}

\author{M. S. Bisht}
\affiliation{Indian Centre for Space Physics, 466 Barakhola, Netai Nagar, Kolkata, 700099, West Bengal, India}

\author{F. M. Walter}
\affiliation{Department of Physics and Astronomy, Stony Brook University, Stony Brook, NY 11794-3800, USA}

\author{R. Pandey}
\affiliation{Physical Research Laboratory, Navrangpura, Ahmedabad, Gujarat 380009, India}

\author{C. E. Woodward}
\affiliation{Minnesota Institute for Astrophysics, University of Minnesota, 116 Church Street SE, Minneapolis, MN 55455, USA}

\author{D. E. Harker}
\affiliation{Department of Astronomy and Astrophysics, University of California, San Diego, 9500 Gilman Drive, La Jolla, CA 92093-0424, USA}

\author{D. Bisht}
\affiliation{Indian Centre for Space Physics, 466 Barakhola, Netai Nagar, Kolkata, 700099, West Bengal, India}

\author{H. P. Singh}
\affiliation{Department of Physics and Astrophysics, University of Delhi, 110007 Delhi, India}

\author{A. Agarwal}
\affiliation{Center for Cosmology and Science Popularization(CCSP) SGT University, Budhera, Delhi- NCR - 122006, India}

\author{J. C. Pandey}
\affiliation{Aryabhatta Research Institute of Observational Sciences (ARIES), Manora Peak, Nainital 263129, India}

\author{A. Joshi}
\affiliation{Institute of Astrophysics, Pontificia Universidad Católica de Chile, Av. Vicuña MacKenna 4860, 7820436 Santiago, Chile}

\author{K. Belwal}
\affiliation{Indian Centre for Space Physics, 466 Barakhola, Netai Nagar, Kolkata, 700099, West Bengal, India}

\author{Christian Buil}
\affiliation{Castanet Tolosan Observatory, 6 place Clemence Isaure, F-31320 Castanet Tolosan, France}

\keywords{stars : novae, cataclysmic variables - stars : individual (V5579 Sgr) - techniques : spectroscopic - line : identification} 

\begin{document}

\begin{abstract}
V5579 Sgr was a fast nova discovered in 2008 April 18.784 UT. We present the optical spectroscopic observations of the nova observed
from the Castanet Tolosan, SMARTS and CTIO observatories spanning over 2008 April 23 to 2015 May 11. The spectra are dominated by hydrogen Balmer, Fe II and O I lines with P-Cygni profiles in the early phase, typical of an Fe II class nova. The spectra show He I and He II lines along with forbidden lines from N, Ar, S, and O in the nebular phase. The nova showed a pronounced dust formation episode that began about 20 days after the outburst. The dust temperature and mass were estimated using the WISE data from spectral energy distribution (SED) fits. The PAH-like features are also seen in the nova ejecta in the mid-IR Gemini spectra taken 522 d after the discovery. Analysis of the light curve indicates values of t$_2$ and t$_3$ about 9 and 13 days, respectively, placing the nova in the category of fast nova. The best fit \textsc{cloudy} model of the early decline phase JHK spectra obtained on 2008 May 3 and the nebular optical spectrum obtained on 2011 June 2 shows a hot white dwarf source with T$_{BB}$ $\sim$ 2.6 $\times$ 10$^5$ K having a luminosity of 9.8 $\times$ 10$^{36}$ ergs s$^{-1}$. Our abundance analysis shows that the ejecta is significantly enhanced relative to solar, O/H = 32.2, C/H = 15.5 and N/H = 40.0 in the early decline phase and O/H = 5.8, He/H = 1.5 and N/H = 22.0 in the nebular phase.
\end{abstract}

\section{Introduction}
Classical novae (CNe) are explosive events characterized by a sudden and substantial increase in brightness of a primary white dwarf (WD) star in a semi-detached binary star system. The WD accretes hydrogen-rich matter from its companion via an accretion disc. 
This mass transfer is made through an inner Lagrange point (L$_1$). The material accumulates on the
surface of the WD, until eventually the matter at the base of the accreted envelope undergoes 
runaway nuclear fusion reactions, releasing an immense amount of energy and causing the luminosity of the underlying binary system to increase upto $10^{4}$ to $10^{5}$ L$_{\odot}$. The luminosity of the quiescence stage generally ranges from a few times to several hundreds times L$_{\odot}$ \citep{BodeEvansBook2008}. Approximately $10^{-4} - 10^{-5}$ M$_{\odot}$ of nuclear-processed material is ejected into the interstellar medium (ISM), typically at velocities ranging from a few hundred to several thousand kilometers per second \citep[][and references therein]{BodeEvansBook2008,Jose2020,Starrfield2020}. CNe also serves as the main contributor of $^{13}$C, $^{15}$N, and $^{17}$O within the Galaxy, potentially influencing the abundance of other isotopes in the intermediate mass range, such as $^{7}$Li and $^{26}$Al \citep{Starrfield2020}. 

Nova V5579 Sgr was discovered during eruption
in 2008 April 18.784 UT by Nishiyama and Kabashima \citep{2008IAUC.8937....1N} at V = 8.4. \citet{munari.1352....1M} reported a rapid and steady brightening of about 0.7 mag/d in the initial stages of V5579 Sgr. The nova reached its maximum brightness of V$_{max}$ = 6.61 on 2008 April 23.394 UT approximately 4.6 days after discovery. A BVRcIc photometric sequence around V5579 Sgr was reported by \citet{heden_murani.5834....1H}. They found that the field is extremely crowded, with several very faint
field stars lying within 4 arcsec of the nova position, and it was not listed in the United States Naval Observatory B1 (USNO B1)
or Two Micron All Sky Survey (2MASS) catalogs.
Astrometry was performed using subprogram Library (SLALIB) \citep{wallace1994slalib} linear plate transformation routines in conjunction with the Second U.S. Naval Observatory CCD Astrograph Catalog (UCAC2). The astrometric fit was less than 0.3 arcsec.

Archival photometry was performed by \citet{jurdana.5839....1J} for the nova. The Asiago Schmidt plate archive was
searched for V5579 Sgr and 106 plates were found that covered its position.
After plate inspection, 58 good B and Ic band plates were finally retained.
These 58 good plates covered the period 1961 June 16 to 1977 July 24, with
an average limiting magnitude B $\sim$ 18, Ic $\sim$ 15.5. They found that the
progenitor was below the limiting magnitude on all plates.
A search of pre-eruption object in the Digitized Sky Survey
(DSS) red image from 1991 and U.K. Schmidt red plate  obtained on 1996 Sept. 08 by \citet{dvorak.1342....2D} did not reveal
any object at the position of V5579 Sgr. With the limiting magnitude of
these surveys close to 20 magnitudes, V5579 Sgr is one of the largest
amplitude ($\Delta$V = 13 magnitudes) observed novae in recent years.  

The optical spectrum taken one day after the discovery by \citet{fuji.8937....1N} showed hydrogen Balmer series absorption lines, with a prominent P Cygni profile in the H$\alpha$ line in addition to several additional broad absorption lines. Infrared (IR) spectra reported by \citet{raj2011nirnova}, \citet{2008IAUC.8948....1R}, and \citet{rudy2008v5579} showed lines of O I, N I, Ca II and strong lines of C I. The full width at half maximum (FWHM) of the lines was estimated as 1600 km~s$^{-1}$. 
The NIR spectrophotometric evolution of the nova from 5 days to 25 days after discovery was presented by \citet{raj2011nirnova}. The JHK band spectra taken around the maximum show prominent lines of hydrogen, neutral nitrogen, and carbon and show deep P-Cygni profiles. Dust formation was reported both by \citet{rudy2008v5579} and \citet{raj2011nirnova}, which is consistent with the presence of spectral lines of low-ionization species such as Na I and Mg I in the early spectra, as they are believed to be indicators of low-temperature zones conducive to dust formation in the nova ejecta. Dust formation was also seen from the light curve as a sudden increase in the rate of
decay in the V-band brightness after about 20 days after the outburst. The \textit{Neil Gehrels Swift Observatory} observed V5579 Sgr during six epochs from 2008 April 28 to 2009 July 31, with no X-ray detection \citep{schwarz2011swift}.
\par
The paper is organized as follows. Section~\ref{observations} presents the details of the observations. Section~\ref{analysis} investigates the evolution of optical spectra from days 4 to 2578 post-outburst. This section also includes a comprehensive analysis of the nova, estimation of crucial physical and chemical parameters using photoionization modeling, along with a discussion on dust using archival low-resolution mid-infrared (IR) Gemini spectra obtained 522 days post-outburst and WISE data. Finally, a comprehensive discussion in Section~\ref{discussion} and summary in Section~\ref{summary} is provided.

\section{Observations}\label{observations}
Low-dispersion optical spectra were obtained using the Castanet Tolosan observatory in France and facilities of the Small and Moderate Aperture Research Telescope System (SMARTS) in Chile.
The six low-dispersion spectra from the Castanet Tolosan observatory were obtained starting from near-peak brightness on 2008 April 23 through 2008 April 29. We obtained 15 spectra with the SMARTS R/C spectrograph on an irregular cadence from 2008 April 27 through 2011 July 28. 
The SMARTS R-C grating spectrograph, data reduction techniques, and observing modes are described by \citet{wal12}. A final low-dispersion spectrum was obtained on 2015 May 11 using the COSMOS
long slit spectrograph on the Cerro Tololo Inter-American Observatory (CTIO\endnote{\url{https://noirlab.edu/science/}}) Blanco 4m telescope. 
Data was reduced using standard IRAF procedures; spectra were extracted using
software written in IDL. An observation of the spectrophotometric standard
LTT 4364 was used to provide flux calibration.

\subsection{Gemini South TReCs}\label{obstrecs}    

Mid-infrared (IR) 10~$\mu$m{} spectra of V5759 Sgr were obtained on September 23, 2009 00:58.33UT (Program GS-2009B-Q68) using the Thermal-Regions Camera Spectrograph \citep[TReCS;][] {2005hris.conf...84D} 
imaging spectrograph on the 8~m Gemini South telescope on Cerro Pachon, Chile. Spectra were obtained with a  $\simeq 0.65$ arcsec wide slit and the 10~$\mu$m{} Lo-Res grating, using standard IR 
chop-nodding observing techniques, under conditions of low relative humidity and measured seeing 
of $\simeq 0.32$ arcsec in the [N]-band (determined from the acquisition images). The V5579~Sgr 
spectra have a total exposure time of 642~s. The photometric standard was HD~151680. Raw
data sets (two sets of spectra) were retrieved from the Gemini Observatory Science Archive
Data\footnote{\url{https://archive.gemini.edu/searchform}} for reduction following the methodology outlined
in \citet{harker2018hyperactivity}. The spectra were flux calibrated using the average [N]-band flux value of 1.77~Jy 
calculated from acquisition image photometry. The spectra were integrated over the width of the [N]-band filter 
and ratioed to the calculated flux value.

The log of spectroscopic observations is given in Table \ref{log}.
\begin{table}
\centering
\caption{Observational log for spectroscopic data obtained for V5579 Sgr.}
\label{log}
\resizebox{\hsize}{!}{%
\begin{tabular}{lcclcc}
\hline
\hline
& \textbf{Time since}  & \textbf{Exposure}   & \textbf{Wavelength}  & \textbf{Observatory} \\
\textbf{Date (UT)} & \textbf{discovery} & \textbf{time} & \textbf{range} &\\
& \textbf{(days)} & \textbf{(s)} &   \textbf{(\AA)} & \\
\hline
2008 April 23.08 & 4.31 & 240  & 4293--6856  & Tolosan \\[0.25ex]
2008 April 24.10 & 5.32 & 2400 & 4288--6860  & Tolosan \\[0.25ex]
2008 April 25.09 & 6.31 & 1500 & 4290--6857  & Tolosan \\[0.25ex]
2008 April 26.06 & 7.27 & 1500 & 4296--6854  & Tolosan \\[0.25ex]
2008 April 27.09 & 8.13 & 2400 & 4722--6865  & Tolosan\\[0.25ex]
2008 April 28.18 & 9.39 & 600  & 4060--4734  & SMARTS \\[0.25ex]
2008 April 28.38 & 9.59 & 900 & 3869--4543 & SMARTS \\[0.25ex]
2008 April 29.10 & 10.32 & 2700  & 5579--7242  & Tolosan \\[0.25ex]
2008 April 29.34 & 10.55 & 1200 & 3870--4544  & SMARTS \\[0.25ex]
2008 May 01.28 & 12.49 & 180  & 5533--6853  & SMARTS \\[0.25ex]
2008 May 03.35 & 14.57 & 1200  & 3871--4545 & SMARTS \\[0.25ex]
2008 May 04.32 & 15.54 & 300  & 3654--5425  & SMARTS \\[0.25ex]
2008 May 10.14 & 21.36 & 1200  & 4061--4735  & SMARTS \\[0.25ex]
2008 May 10.27 & 21.49 & 300  & 5528--6847   & SMARTS \\[0.25ex]
2008 May 11.25 & 22.47 & 300  & 3655--5426  & SMARTS \\[0.25ex]
2008 May 16.32 & 27.53 & 900  & 3654--5424  & SMARTS \\[0.25ex]
2009 Sept 23.04 & 522.26 & 642  & 77000--140000 & Gemini \\[0.25ex]
2011 May 17.37 & 1123.58 & 2700  & 5631--6950 & SMARTS \\[0.25ex]
2011 May 18.16 & 1124.38 & 1800  & 5987--9482 & SMARTS \\[0.25ex]
2011 June 02.17 & 1139.39 & 2700  & 2760-9600  & SMARTS \\[0.25ex]
2011 June 03.35 & 1140.57 & 2700  & 5625--6945 & SMARTS \\[0.25ex]
2011 July 28.22 & 1195.43 & 3000  & 5996--9490 & SMARTS \\[0.25ex]
2015 May 11.20 & 2578.41 & 2700 & 3800-6615 & CTIO \\[0.25ex]
\hline
\end{tabular}}
\end{table}
 
\section{Analysis}\label{analysis}
\subsection{The pre-maximum rise, outburst luminosity, reddening and distance}\label{ outburst luminosity, reddening and distance}

\begin{figure}
 \includegraphics[width=1\columnwidth]{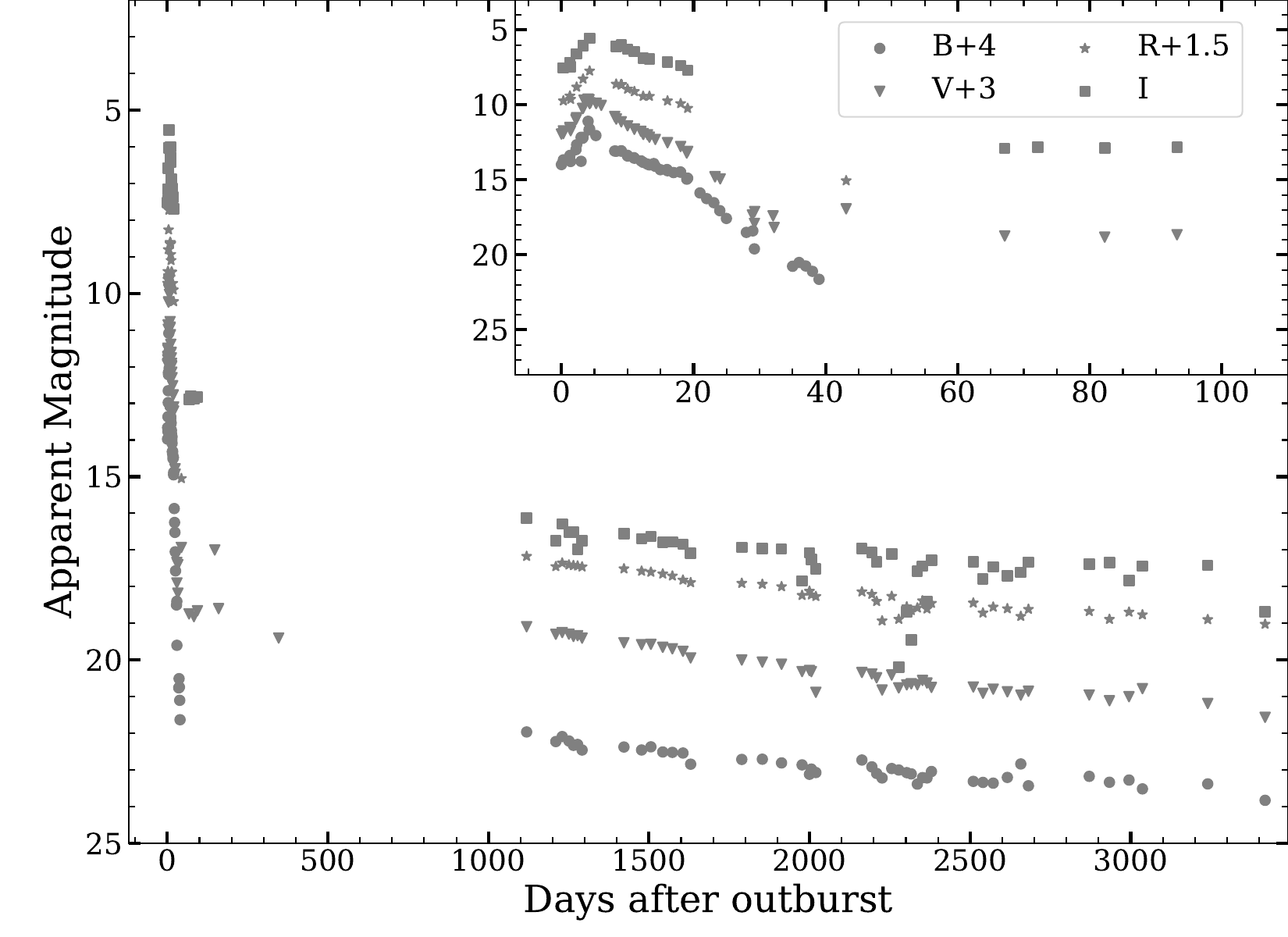}
 \caption{Optical light curves of V5579 Sgr generated using optical data from AAVSO and SMARTS. Offsets have been applied for all the magnitudes except $I$ for clarity.}
 \label{lc_optical}
\end{figure}

\begin{figure}
 \includegraphics[width=0.9\columnwidth]{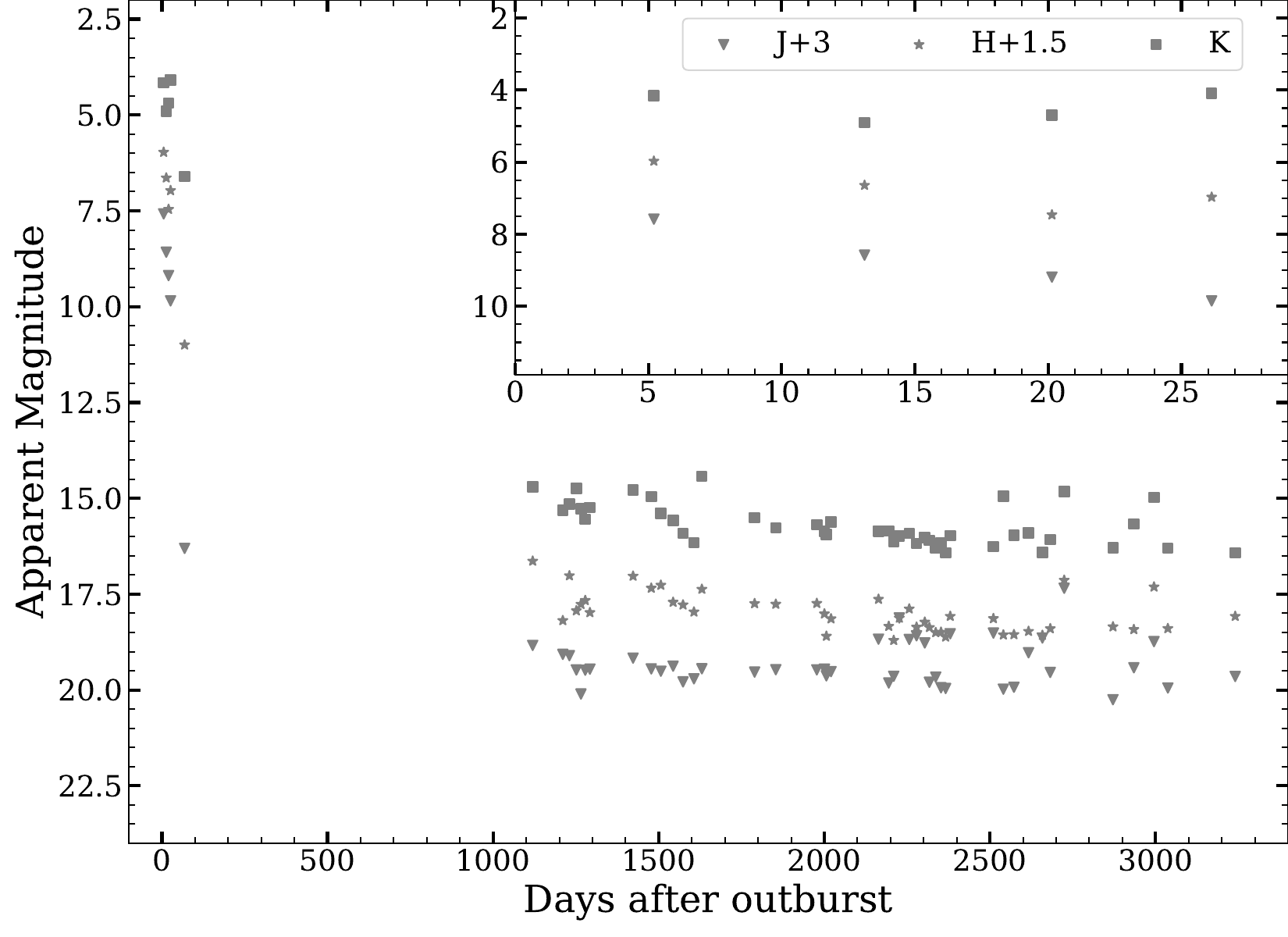}
 \caption{NIR light curves of V5579 Sgr generated using NIR data from \citet{raj2011nirnova} and SMARTS. Offset have been applied for all the magnitudes except $K$ for clarity.}
 \label{lc_nir}
\end{figure}

The optical and near-IR (NIR) light curves based on the American Association of Variable Star Observers (AAVSO\endnote{\url{https://www.aavso.org/}}) database \citep{kafka2021observations}, \citet{raj2011nirnova} and the SMARTS Nova Atlas\endnote{\url{http://www.astro.sunysb.edu/fwalter/SMARTS/NovaAtlas/}} \citep{wal12} database are presented in Figs \ref{lc_optical} and \ref{lc_nir}.
Nova V5579 Sgr was first discovered on 2008
April 18.784 UT; in this paper, we considered this date as the day of the outburst (day 0).
The BVRI band light curve begins within a day after discovery. The brightness increased at a fast rate for all bands and reached the peak magnitude V$_{max}$ = 6.61 $\pm$ 0.01 after 4.6 days of the discovery. The brightness in the BVRI bands shows a moderate decline after the peak until day 18. A sudden increase in the decay rate was observed in the $B$ and $V$ bands after 19.6 days. The decay rate was about 0.22 $\pm$ 0.03 mag/day, until 19.6 days, and changed to 0.39$\pm$ 0.04 mag/day after 20 days. This clearly indicates dust formation in the nova ejecta (see Figure \ref{lc_optical}). This is further supported by Figure \ref{lc_nir}, which shows an increase in the brightness of the NIR $H$ and $K$ bands.
The nova faded to >15 mag in $V$ between 32 and 67 days from the outburst, suggesting that a large amount of dust was formed in the nova ejecta. We do not have any data between 350-1125 days, so we are not able to comment on recovery in the BVRI and JHK bands. Since day 1125 it has faded slowly, in optical and NIR which continued until our last observation at day 3419. The nova was at 19.8 $\pm$0.002 , 18.6 $\pm$0.001, 17.5 $\pm$0.001, 18.7 $\pm$0.001 and 15.1 $\pm$0.01, 17.1 $\pm$0.001, 16.2 $\pm$0.001 mag in BVRI bands and JHK bands, respectively on day 3419. 

\begin{figure*}
\begin{center}
\includegraphics[width=1\columnwidth]{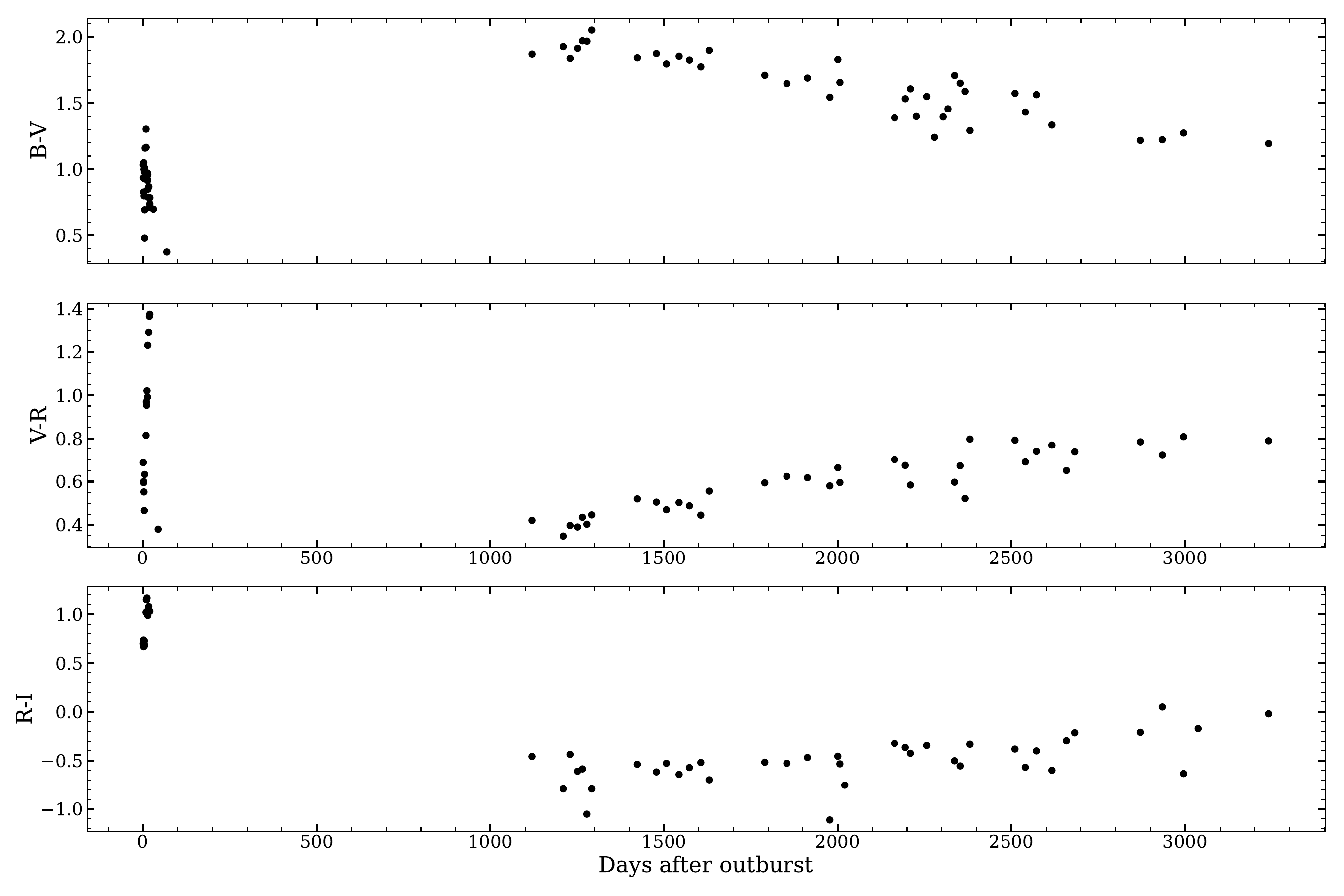}
	\caption{Evolution of Optical color terms of nova V5579 Sgr from day 1 to day 3240 since discovery.}
	\label{optical_color}
\end{center}
\end{figure*}

\begin{figure*}
\begin{center}
\includegraphics[width=1\columnwidth]{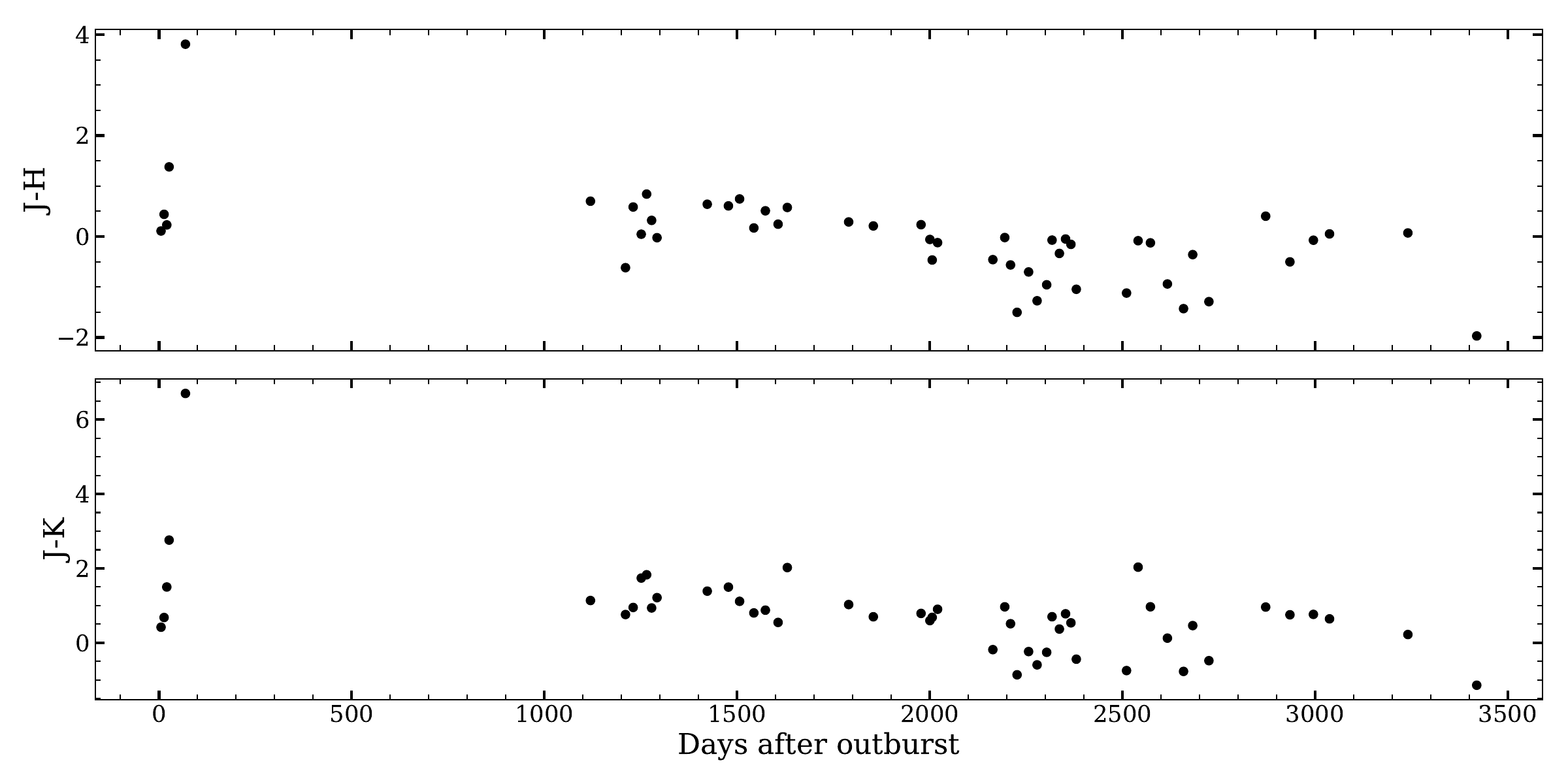}
	\caption{Evolution of NIR color terms of nova V5579 Sgr from day 1 to day 3419 since discovery.}
	\label{jhk_color}
\end{center}
\end{figure*}

We derive the reddening to be E(B - V) = 0.7 $\pm$0.14 from the mean value of all reported values (0.72, 0.82 and 0.56; \cite{raj2011nirnova}, \cite{hachisu2019light}, \cite{hachisu2021ubv}, respectively) towards the direction of the nova. We adopt A$_V$ = 2.2 mag for extinction, assuming R = 3.1. 

From the least-squares regression fit to the post-maximum V band light curve, we estimated t$_2$ to be 9 $\pm$ 0.2 days. Using the maximum magnitude versus rate of decline (MMRD) relation of \citet{downes2000mmrd}, we would expect that the absolute magnitude of the nova
is M$_V$ = -9.3 $\pm$ 0.1 giving a distance modulus of V$_{max}$ - M$_V$ = 15.9 mag. Using the distance modulus relation, we obtain a value of the distance d = 5.6 $\pm$ 0.2 kpc to the nova. These estimated values of M$_V$ and the distance are different from the estimates of \citet{raj2011nirnova} mainly due to the different MMRD relation they used and the slight variation in the estimated value of t$_2$.

These observed values of the amplitude $\Delta$V = 13 and t$_2$ $\sim$ 9 days for V5579
Sgr are consistent with its location in the amplitude versus decline rate plot
for CNe presented by \citet{warner2008properties}, which shows $\Delta$ V = 12 - 15 for
t$_2$ = 9 days. 

We have estimated the mass of the white dwarf ($M_{WD}$) using the derived value of the
absolute magnitude M$_V$, t$_3$ from the relations given by \citet{livio1992classical}.
The mass of the underlying white dwarf in V5579 Sgr is estimated
to be 1.16 M$_{\odot}$. The outburst luminosity is estimated from $M_{bol}$ = 4.8 + 2.5log (L/$L_{\odot}$), where the bolometric correction applied to $M_V$ is assumed to lie between -0.4 and 0.00 corresponding from A to F spectral types, respectively (novae at maximum generally have a spectral type between A to F). Using $M_V$ = -9.3, we calculate the luminosity of the outburst (5.34 $\pm$ 0.97)$\times$ $10^5$ $L_{\odot}$. 
A classification system for the optical light curves for novae on the basis of the
shape of the light curve and the time to decline by 3 magnitudes ($t_3$) from V$_{max}$ has been
presented by \citet{strope2010catalog}. The shape of the optical
light curve of V5579 Sgr presented in Figure \ref{lc_optical} has all the characteristics
of the D class of nova, which shows a dust dip in the V band light curve after
the optical maximum. The early decline of V5579 Sgr following the rise to the
maximum is interrupted by fast decline around 15 days after optical maximum
and continues further. Thus, the classification of the optical light curve for
V5579 Sgr is D(13), since the estimated value of $t_3$ is $\sim$ 13 days for V5579 Sgr.
The value $t_3$ obtained using the AAVSO V band light curve is slightly lower than the value obtained by the relation $t_3$ = 2.75 ($t_2$)$^{0.88}$, which gives $t_3$ = 18.8
 days \citep{1995cvs..book.....W}.
The small observed value of t$_2$ makes V5579 Sgr as one of the fastest Fe II novae in recent years. 
Other fast novae of Fe II class in recent years include N Aql 1999 (V1494 Aql, t$_2$
= 6.6 d, \cite{2000A&A...355L...9K}), N Sgr 2004 (V5114 Sgr, t$_2$ = 11 d, \cite{ederoclite2006}), N Cyg 2005 (V2361 Cyg, t$_2$ =
6 d, \cite{hachisu2007universal}), N Cyg 2006 (V2362 Cyg, t$_2$ = 10.4 d, \cite{munari2008nature}), N Oph 2006 (V2576 Oph, t$_2$ =
8 d, \cite{schwarz2011swift}), N Cyg 2007 (V2467 Cyg, t$_2$ = 7 d, \cite{schwarz2011swift}), and N Cyg 2008 (V2468 Cyg, t$_2$ =
7.8 d,  \cite{iijima2011spectral}). 

The NIR light curves are made using data from \citet{raj2011nirnova} and the SMARTS/CTIO 1.3 m telescope facility \citep{wal12}. The NIR light curves are non-monotonic (Fig. \ref{lc_nir}). The J band light curves showed a steep, steady decline in the initial phase from the outburst, lagging the dust dip seen in the optical after day 20. But the HK band light curves showed an increase after 20 days from outburst, which is consistent with the dust formation as the dust mainly contributes at longer wavelengths \citep[and the emission lines were not extremely strong][]{rudy2008v5579}. The K band light curve might have reached a peak value between 20 and 70 days after the outburst. After about day 20 the K band brightness increased by over 0.3 mag, accompanied by a smaller brightening in H, while J continued to fade. This can be taken as the start of dust formation. The K-band magnitude brightened by 0.5 mag on day 26. 

The B-V, V-R, and R-I color evolution is shown in Fig. \ref{optical_color}. The B-V, V-R, and R-I colors were approximately 1.03, 0.68, and 0.70, respectively, on day 1. By day 20, the B-V color decreased to 0.74, while V-R and R-I increased to 1.36 and 1.03, respectively. Following this, B-V continued to decrease to 0.37 by day 68, and V-R decreased to 0.38 by day 43. There were no observations between days 68 and 1119. On day 1119, the B-V, V-R, and R-I colors were approximately 1.87, 0.42, and -0.46, respectively. From day 1119 to day 3240, B-V slowly decreased to 1.19, while V-R and R-I slowly increased to 0.79 and -0.02 mag, respectively.
The J-K and J-H color evolution is shown in Fig. \ref{jhk_color}. The J-K and J-H colors were about 0.4 and 0.1 at peak brightness, respectively. No color excess was seen until 8 May 2008 (20.1 days since discovery), but a constant increase in the colors was clearly visible. The J-K, J-H and H-K colors reach a maximum value of 6.7, 3.8 and 2.8, respectively after 68 days from discovery which indicate that dust has been formed in the nova ejecta.

\subsection{Line identification, general characteristics and evolution of the optical spectra}
\begin{figure*}
\centering
\includegraphics[width=1.0\linewidth]{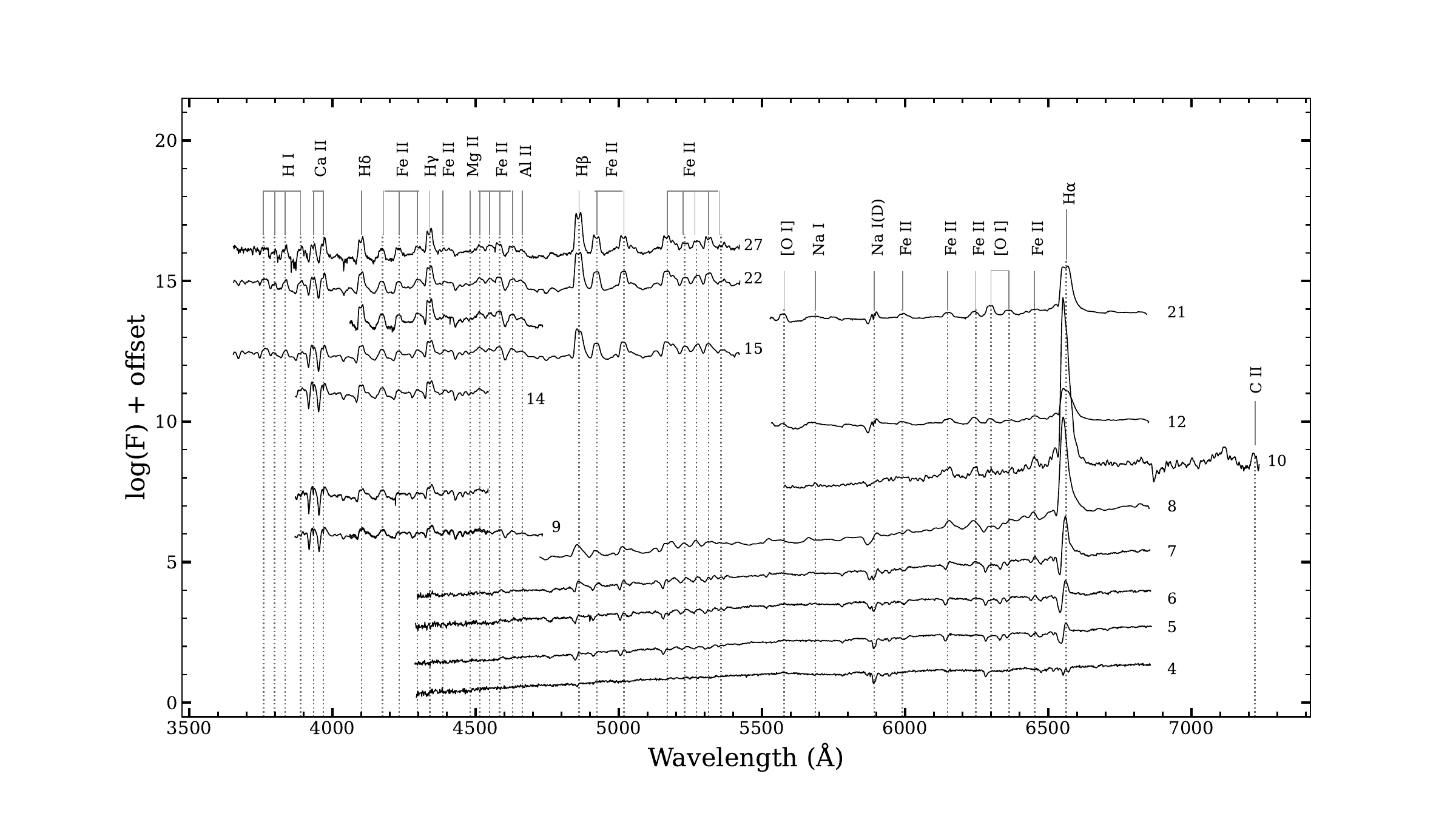}
\caption{Low-resolution optical spectral evolution of V5579 Sgr obtained from day 4 (2008 Apr 23) to day 27 (2008 May 20). Spectra are dominated by Fe II multiplets and hydrogen Balmer lines. The lines identified are marked, and time since discovery (in days) is marked against each spectrum.}
	\label{spectra_early}
\end{figure*}

\begin{figure*}
\centering
\includegraphics[width=1\linewidth]{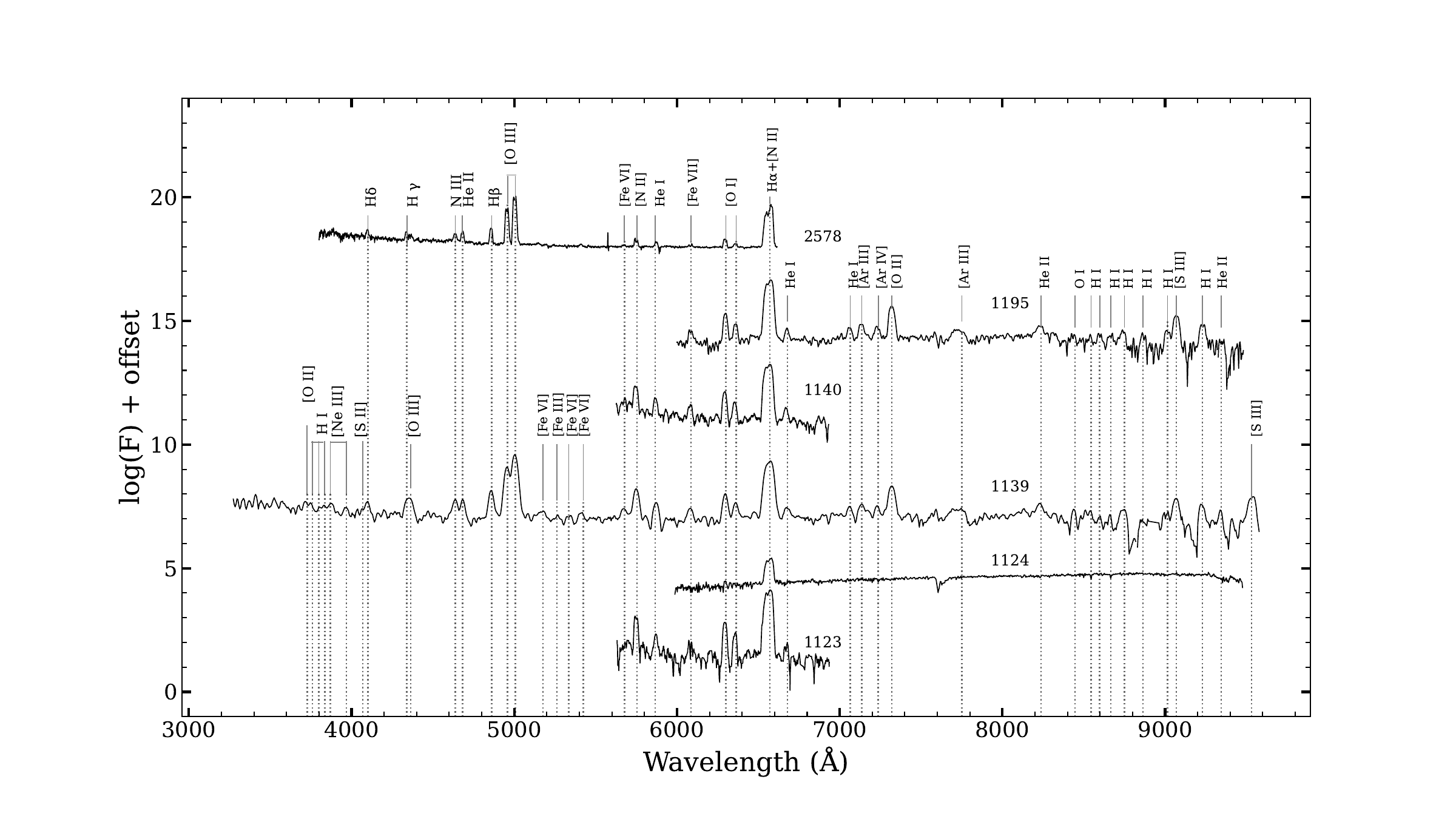}
\caption{Low-resolution optical spectral evolution of V5579 Sgr obtained from day 1123 (2011 May 17) to day 2578 (2015 May 11). The [O I] lines at 6300 and 6363 \AA\ are very prominent. Line identifications are marked, and time since discovery (in days) is marked against each spectrum.}
	\label{spectra_nebular}
\end{figure*}

The optical spectra from SMARTS and the Castanet Tolosan Observatory, presented in Figs. \ref{spectra_early} and \ref{spectra_nebular}, cover the early decline phase with 16 spectra and the nebular phase with five spectra. We presented
15 low-dispersion spectra using the SMARTS 1.5 m/RC spectrograph from 2008 April 28 (day 9) through 2011 July 27 ( day 1195), one low-dispersion spectrum obtained with the CTIO Blanco 4m COSMOS spectrograph on 2015 May 11
(day 2578), and 6 low-dispersion spectra from Castanet Tolosan starting from peak brightness 2008 April 23 (day 4) to 2008 April 29 (day 10). 
The spectrum taken around peak brightness between 4-8 days shows deep P-Cygni profiles for the H$\alpha$ and H$\beta$ lines. The Na II doublets at 5890 and 5896 \AA{} are also seen. The presence of large number of Fe II lines (5018, 5169, 5235, 5276, 6148, 6456 \AA{}) in the initial phase indicates that the nova belongs to Fe II class. Most of the lines are seen in emission on day 9 e.g. H I 4102, 4340 and Fe II 4169, 4233, 4417, 4458 \AA{}. The spectra show the presence of hydrogen Balmer lines and Fe II multiplets along with Ca II (H and K) and He I at 5876, [S II] at 6716 \AA{}. The ejecta velocity was estimated around 1400- 1500 \kms for hydrogen lines. There was no significant change in the optical spectra until our last observations on day 27.

The emission lines grew progressively stronger and more complex with time. The Hydrogen Balmer and other metal lines were initially flat-topped, caused by the high line optical depth and later they developed `double-horn' structure indicating that the total observed emission at a given wavelength arises from material receding from and approaching the observer (i.e., a ring or shell structure to the ejecta). The line profiles evolve from P-Cygni in the pre-maximum phase to a
boxy and structured one in the decline phase. On day 27 for H$\alpha$ line the velocity peak of the blue-shifted material is of the order of $\sim$ -600 $\pm$ 30 \kms, and the red-shifted material by $\sim$ 320 $\pm$ 30 \kms. In the nebular phase (day 1123) it changes to $\sim$ -560 $\pm$ 20 \kms for the blue-shifted and $\sim$ 670 $\pm$ 20 \kms for the red-shifted. In our last observations, the blue-shifted component decreased to -470 $\pm$ 30 \kms the red-shifted to 590 $\pm$ 30 \kms. Although we cannot ignore the contribution of the [N II] line at 6584 \AA{}.

The nebular phase spectra obtained after about 3 years, on day 1123, show higher excitation lines of [N II] 5755, He I 5876, [O I] 6300, 6363 and H$\alpha$ blended with [N II] 6548 and [N II] 6584 \AA{}. The subsequent spectra taken on days 1139 and 1140 show emission lines of [O III] 4363, 4959, 5007, [N II] 5755, He II 4686, [O I] 6300, 6363, He I 5876, [O II] 7320 and [S III] 9069, 9229 \AA{}. In the final spectra the lines of [Ar III] 7136, [Ar IV] 7237 and He II 8237 \AA{} were present. The H$\alpha$ +[N II] lines are not clearly resolved, and [N II] 6584 \AA{} dominates the blend. The nova evolve in PfeAo sequence as per classification given by \citet{williams1991evolution}. 
The only permitted lines are H$\beta$ and H$\alpha$ and some He I and He II
lines. Overall the spectrum resembles that of V2676 Oph some
three years after its outburst \citep{raj2017v2676}.

The day 2578 spectrum is still nebular, and is more
excited. We see N III 4640, He II 4686 and 5411, [Fe VII] 6087, and a lot
of other Nitrogen lines (5460, 5679 in addition to 5755 and 6548/6584 \AA).

\begin{figure}
	\includegraphics[width=\columnwidth]{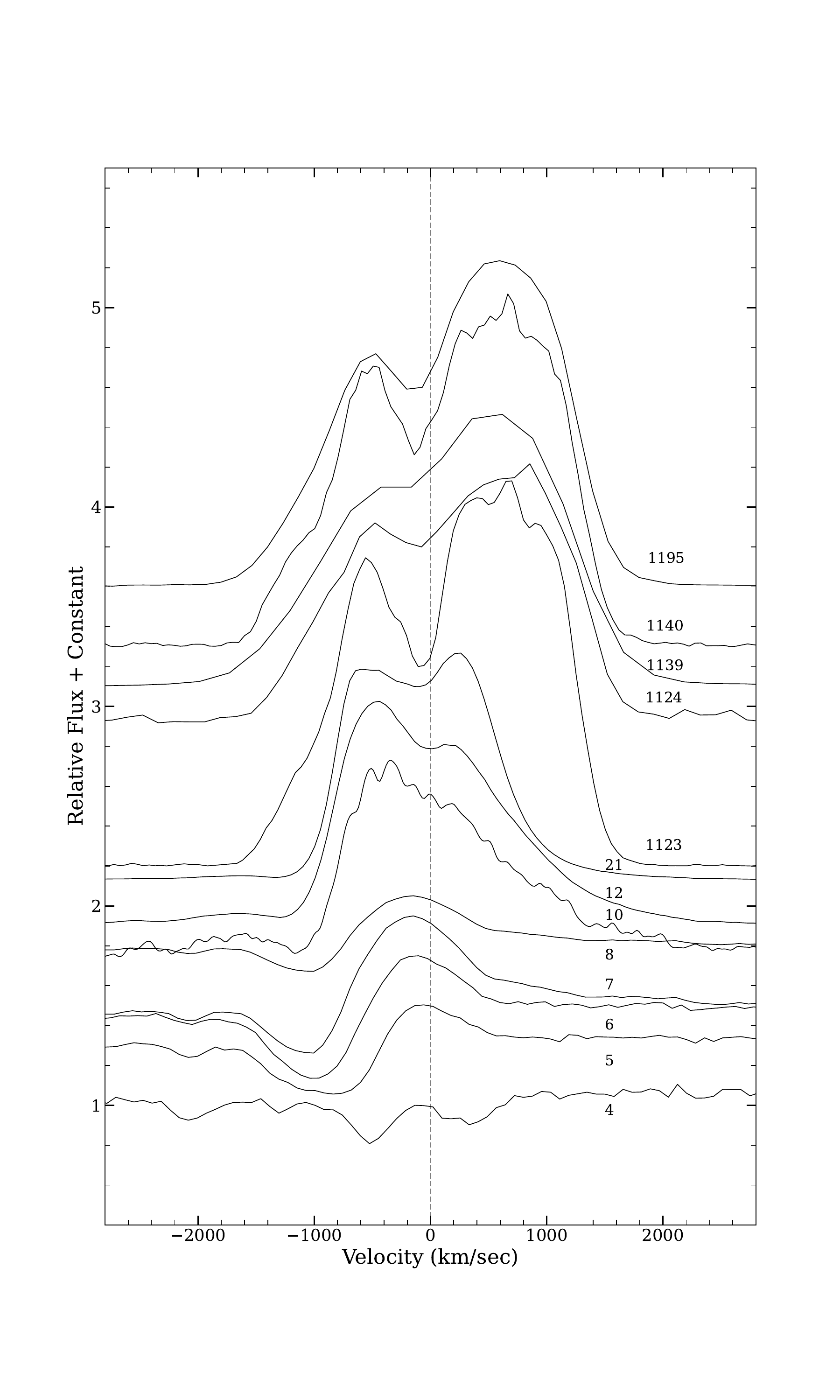}
	\caption{Evolution of H$\alpha$ velocity profile of V5579 Sgr from day 4 (2008 Apr 23) to 2578 (2011 July 28) obtained using the low resolution spectroscopic data. The line profiles evolve from P-Cygni in the pre-maximum phase to a boxy and structured one in the decline phase.}
	\label{halpha}
\end{figure}

The evolution of the H$\alpha$ velocity profile is as shown in Fig. \ref{halpha}. The P-Cygni profile was present until day 10.
The P-Cygni profile of the H$\alpha$ line has a blue-shifted component at $\sim$ - 500 \kms on day 4, after which the bulk of the absorption shifted to $\sim$ - 1000 \kms.

\subsection{Physical parameters}
The physical parameters e.g., optical depth, electron temperature, density, and oxygen mass, can be estimated from the dereddened line fluxes of oxygen from the optical spectrum. 

The optical depth of oxygen and the temperature of the electron are estimated using the formulas found in \citet{williams1994extinction}, 
\begin{align}
\label{op_dep}
\dfrac{F_{\lambda6300}}{F_{\lambda6364}} = \dfrac{(1 - e^{-\tau})}{(1-e^{-\tau/3})}.
\end{align}
Using the value of $\tau$, the electron temperature (K) is given by
\begin{align}
\label{Te}
T_e =   \dfrac{11200}{ log [\dfrac{(43\tau)}{(1-e^{-\tau})} \times \dfrac{F_{\lambda6300}}{F_{\lambda5577}}]} 
\end{align}
where F$_{\lambda5577}$, F$_{\lambda6300}$ and F$_{\lambda6364}$ are the line intensities of [O I] 5577, 6300 and 6364 \AA\ lines. The optical depth $\tau$ of the ejecta for [O I] 6300 \AA{} is estimated to be about 0.23 on days 21 and 1139. We do not see any change in the optical depth in three years of evolution. The electron temperature was estimated to be around 5700 K for day 21 which is typically seen in novae (\cite{ederoclite2006}, \cite{williams1994extinction}). 
We have also estimated the electron density on day 1139 (taking the typical value of T$_e$ $\sim$ 10$^4$K) using the [O III] 4363, 4959 and 5007 \AA{} lines in the following relation from \citet{ost06} 
\begin{equation}
\frac{j_{4959}+j_{5007}}{j_{4363}} = 7.9 \frac{e^{3.29\times10^{4}/T_{e}}}{1+4.5\times10^{-4}\frac{N_{e}}{T_{e}^{1/2}}}
\end{equation}
In our low resolution spectra [O III] 4363 line was blended with $H\gamma$ 4340 \AA\ , using the IRAF\endnote{NOIRLab IRAF is distributed by the Community Science and Data Center at NSF NOIRLab, which is managed by the Association of Universities for Research in Astronomy (AURA) under a cooperative agreement with the U.S. National Science Foundation.} SPLOT deblending tool, the emission lines were deblended, and the flux for [O III] 4363 \AA\ was calculated. The electron density Ne was estimated as 5.42 $\times$ 10$^{5}$ cm$^{-3}$.
 The oxygen mass can be estimated using the relation of \citet{williams1994extinction},
 \begin{equation}
\dfrac{m(O)}{M_{\odot}} = 152 d_{kpc}^{2} exp[\dfrac{22850}{T_e}] \times 10^{1.05E(B-V)}\frac{\tau}{1 - e^{-\tau}} F_{\lambda6300}
\end{equation}

The oxygen mass thus determined is found to be 3.8 $\times$ 10$^{-7}$ M$_{\odot}$.
Here we have used the distance estimated in Section \ref{ outburst luminosity, reddening and distance}. 

\begin{table}
\caption{Observed and best-fit NIR \textsc{cloudy} model line flux ratios for day 15 of V5579 Sgr }
\label{nir_table}
\begin{center}
\resizebox{\hsize}{!}{%
\begin{tabular}{lclcc}
\hline
\hline
\textbf{Line ID} & \boldmath{$\lambda$ ($um$)} & \textbf{Observed}$^{a}$ & \textbf{Modelled}$^{a}$ & \boldmath{$\chi^2$}  \\
                    \hline
\  &  & J-band &  &  \\  
                     \hline
                Pa$\gamma$           & 1.0938     & 7.82E$-$01 & 5.66E$-$01 & 7.27E$-$01 \\
                O I + C I          & 1.1295     & 1.88E$+$00 & 1.40E$+$00 & 2.58E$+$00 \\
                C I           & 1.1674     & 3.63E$-$01 & 1.32E$-$01 & 5.90E$-$01 \\
                C I           & 1.1755     & 6.99E$-$01 & 1.20E$-$01 & 3.72E$+$00 \\
                C I           & 1.1895     & 3.77E$-$02 & 6.71E$-$02 & 9.66E$-$03 \\
N I           & 1.2469     & 1.52E$-$01 & 1.01E$-$02 & 2.20E$-$01 \\
Blnd C I      & 1.2575     & 2.54E$-$01 & 1.60E$-$01 & 1.42E$-$01 \\
Pa $\beta$          & 1.2818     & 1.00E$+$00 & 1.00E$+$00 & 0.00E$+$00 \\
C I           & 1.2970     & 1.96E$-$02 & 5.50E$-$01 & 3.12E$+$00 \\
                \hline
\  &  & H-band &  &  \\
\hline
 Br 19         & 1.5256     & 2.22E$-$01 & 5.57E$-$01 & 1.24E$+$00 \\
Br 18         & 1.5341     & 3.00E$-$01 & 5.95E$-$01 & 1.39E$+$00 \\
Br 17         & 1.5439     & 2.80E$-$01 & 6.39E$-$01 & 1.42E$+$00 \\
   Br 16         & 1.5557     & 4.85E$-$01 & 6.88E$-$01 & 6.63E$-$01 \\
 Br 15         & 1.5685     & 5.51E$-$01 & 7.46E$-$01 & 9.46E$-$01 \\
                Br 14         & 1.5881     & 6.95E$-$01 & 8.14E$-$01 & 2.27E$-$01 \\
                C I           & 1.6005     & 2.66E$-$01 & 1.92E$-$01 & 8.54E$-$02 \\
                Br 13         & 1.6109     & 5.91E$-$01 & 8.96E$-$01 & 1.04E$+$00 \\
                Br 12         & 1.6407     & 1.00E$+$00 & 1.00E$+$00 & 0.00E$+$00 \\
                Br 11 + C I   & 1.6865     & 2.33E$+$00 & 2.83E$+$00 & 2.69E$+$00 \\
                C I           & 1.7045     & 4.48E$-$01 & 1.43E$-$01 & 1.03E$+$00 \\
                Br 10         & 1.7362     & 1.38E$+$00 & 1.31E$+$00 & 4.56E$-$02 \\
                C I           & 1.7448     & 9.48E$-$01 & 4.37E$-$01 & 2.91E$+$00 \\
                C I           & 1.7672     & 8.72E$-$01 & 9.18E$-$01 & 2.36E$-$02 \\
                C I           & 1.7790     & 1.48E$+$00 & 6.14E$-$01 & 8.38E$+$00 \\
                C I          & 1.8022      & 7.92E$-$01 & 5.11E$-$01 & 8.83E$-$01 \\
                \hline
\  &  & K-band &  &  \\
\hline
                C I           & 2.1023     & 1.75E$-$01 & 3.50E$-$01 & 4.96E$-$01 \\
                Blnd C I     & 2.1265     & 4.29E$-$01 & 8.58E$-$02 & 1.31E$+$00 \\
                Br $\gamma$           & 2.1655     & 1.00E$+$00 & 1.00E$+$00 & 0.00E$+$00 \\
                Na I          & 2.2056     & 6.80E$-$02 & 2.52E$-$02 & 2.04E$-$02 \\
                C I           & 2.2156     & 2.09E$-$01 & 3.95E$-$01 & 3.89E$-$01 \\
                C I           & 2.2906     & 1.84E$-$01 & 4.08E$-$01 & 5.58E$-$01 \\
\hline
 \end{tabular}}
\end{center}
\end{table}

\begin{table}
\caption{Best-fit NIR \textsc{cloudy} model parameters obtained on day 15 for the system V5579 Sgr.}
\label{nir_parameters}
\begin{center}
\resizebox{\hsize}{!}{%
\begin{tabular}{l c }
\hline\hline
\textbf{Parameter} & \textbf{Day 15} \\ [0.5ex]
\hline
T$_{BB}$ ($\times$ 10$^{4}$ K) & 1.58 $\pm$ 0.05 \\ [0.25ex]
Luminosity ($\times$ 10$^{36}$ erg/s) & 7.76 $\pm$ 0.8 \\ [0.25ex]
Clump Hydrogen density ($\times$ 10$^{11}$ cm$^{-3}$) & 2.82  \\ [0.25ex]
Diffuse Hydrogen density ($\times$ 10$^{9}$ cm$^{-3}$) & 3.55 \\ [0.25ex]
Covering factor (clump) & 0.60 \\ [0.25ex]
Covering factor (diffuse) & 0.40 \\ [0.25ex]
$\alpha$ & -3.00 \\ [0.25ex]
Inner radius ($\times$ 10$^{14}$ cm) & 1.10 \\ [0.25ex]
Outer radius ($\times$ 10$^{14}$ cm) & 2.21 \\ [0.25ex]
Filling factor  & 0.1 \\ [0.25ex]
C/C$_{\odot}$ & 15.5 $\pm$ 3.0 (15)$^{a}$ \\ [0.25ex]
                N/N$_{\odot}$ & 40.00 $\pm$ 5 (1) \\ [0.25ex]
O/O$_{\odot}$ & 32.2 $\pm$ 3.5 (1) \\ [0.25ex]
Ejected mass ($\times$ 10$^{-4}$ M$_{\odot}$) & 1.67 \\ [0.25ex]
Number of observed lines (n) & 31 \\ [0.25ex]
Number of free parameters (n$_{p}$) & 9 \\ [0.25ex]
Degrees of freedom ($\nu$) & 22 \\ [0.25ex]
Total $\chi^{2}$ & 36.9 \\ [0.25ex]
$\chi^{2}_{red}$ & 1.67 \\ [0.25ex]
\hline
\end{tabular}}
\end{center}
$^{a}$The number of lines available to obtain an abundance estimate is as shown in the parenthesis.
\end{table}

\begin{table}
	\caption{Best-fit optical \textsc{cloudy} model parameters obtained on day 1139 for the system V5579 Sgr}
	\label{optical_parameters}
	\begin{center}
		\resizebox{\hsize}{!}{%
			\begin{tabular}{l c } 
				\hline\hline
				\textbf{Parameter} & \textbf{Day 1139} \\ [0.5ex] 
				\hline 
				T$_{BB}$ ($\times$ 10$^{5}$ K) & 2.55 $\pm$ 0.10 \\ [0.25ex]
				Luminosity ($\times$ 10$^{36}$ erg/s) & 9.77 $\pm$ 0.08 \\ [0.25ex]
				Clump Hydrogen density ($\times$ 10$^{6}$ cm$^{-3}$) & 5.63  \\ [0.25ex]
				Diffuse Hydrogen density ($\times$ 10$^{6}$ cm$^{-3}$) & 2.01 \\ [0.25ex]
				Covering factor (clump) & 0.25 \\ [0.25ex]
				Covering factor (diffuse) & 0.75 \\ [0.25ex]
				$\alpha$ & -3.00 \\ [0.25ex]
				Inner radius ($\times$ 10$^{15}$ cm) & 8.91 \\ [0.25ex]
				Outer radius ($\times$ 10$^{16}$ cm) & 1.60 \\ [0.25ex]
				Filling factor  & 0.05 \\ [0.25ex]
				He/He$_{\odot}$ & 1.45 $\pm$ 0.20 (8)$^{a}$ \\ [0.25ex]
                N/N$_{\odot}$ & 22.00 $\pm$ 0.30 (2) \\ [0.25ex]
				O/O$_{\odot}$ & 5.8 $\pm$ 0.8 (7) \\ [0.25ex]
				Ejected mass ($\times$ 10$^{-4}$ M$_{\odot}$) & 6.36 \\ [0.25ex]
				Number of observed lines (n) & 39 \\ [0.25ex]
				Number of free parameters (n$_{p}$) & 10 \\ [0.25ex]
				Degrees of freedom ($\nu$) & 29 \\ [0.25ex]
				Total $\chi^{2}$ & 39.47 \\ [0.25ex]
				$\chi^{2}_{red}$ & 1.36 \\ [0.25ex]
				\hline
		\end{tabular}}
	\end{center}
	$^{a}$The number of lines available to obtain abundance estimate is as shown in the parenthesis.
\end{table}

\begin{table}
	\caption{Observed and best-fit optical \textsc{cloudy} model line flux ratios for day 1139 of V5579 Sgr}
	\label{optical_table}
	\begin{center}
		\resizebox{\hsize}{!}{%
			\begin{tabular}{lclcc}
				\hline
				\hline
				\textbf{Line ID} & \boldmath{$\lambda$ (\AA)} & \textbf{Observed}$^{a}$ & \textbf{Modelled}$^{a}$ & \boldmath{$\chi^2$}  \\
				\hline
                {[}O II{]}    & 3727       & 4.22E$-$01 & 3.53E$-$02 & 1.66E$+$00 \\
                H I           & 3750       & 2.80E$-$01 & 3.87E$-$02 & 6.41E$-$01 \\
                H I           & 3798       & 1.10E$-$01 & 5.88E$-$02 & 2.97E$-$02 \\
				H I           & 3835       & 8.90E$-$02 & 7.90E$-$02 & 1.08E$-$03 \\
				{[}Ne III{]}  & 3869       & 3.51E$-$01 & 8.35E$-$01 & 2.61E$+$00 \\
				{[}Ne III{]}  & 3968       & 3.12E$-$01 & 2.54E$-$01 & 5.47E$-$02 \\
				{[}S II{]}    & 4069       & 9.86E$-$02 & 9.68E$-$02 & 3.70E$-$05 \\
				H I           & 4102       & 4.19E$-$01 & 2.63E$-$01 & 2.72E$-$01 \\
                {[}O III{]}   & 4363       & 1.28E$+$00 & 2.73E$+$00 & 23.51E$+$00 \\
				  N III + {[}Fe III{]} & 4640       & 3.76E$-$01 & 1.36E$-$01 & 2.66E$+$00 \\
				He II         & 4686       & 3.64E$-$01 & 4.48E$-$01 & 1.11E$-$01 \\
				H I           & 4861       & 1.00E$+$00 & 1.00E$+$00 & 0.00E$+$00 \\
				  {[}O III{]}   & 4959       & 9.42E$+$00 & 9.83E$+$00 & 2.63E$+$00 \\
				  {[}O III{]}   & 5007       & 29.56E$+$00 & 29.31E$+$00 & 9.60E$-$01 \\
                {[}Fe VI{]}   & 5176       & 8.40E$-$02 & 1.80E$-$01 & 1.02E$-$01 \\
                {[}Fe III{]}  & 5270       & 3.92E$-$02 & 9.06E$-$02 & 2.94E$-$02 \\
                He II         & 5412       & 8.12E$-$02 & 3.65E$-$02 & 2.22E$-$02 \\
                {[}Fe VI{]}   & 5677       & 1.03E$-$01 & 3.88E$-$02 & 4.54E$-$02 \\
				{[}N II{]}    & 5755       & 8.64E$-$01 & 7.02E$-$01 & 4.21E$-$01 \\
				He I          & 5876       & 2.58E$-$01 & 4.30E$-$02 & 5.14E$-$01 \\
                {[}Fe VII{]}  & 6086       & 1.15E$-$01 & 3.83E$-$02 & 6.66E$-$02 \\
				{[}O I{]}     & 6300       & 3.78E$-$01 & 3.54E$-$01 & 9.86E$-$03 \\
				{[}O I{]}     & 6364       & 1.55E$-$01 & 1.13E$-$01 & 2.79E$-$02 \\
				H I + {[}N II{]} & 6572    & 8.86E$+$00 & 3.46E$+$00 & 323.29E$+$00 \\
                He I          & 6678       & 1.11E$-$01 & 3.02E$-$02 & 7.21E$-$02 \\ 
				He I          & 7065       & 6.10E$-$02 & 4.470E$-$02 & 2.18E$-$03 \\
				{[}Ar III{]}  & 7136       & 6.90E$-$02 & 1.90E$-$01 & 1.65E$-$01 \\
                {[}Ar IV{]}   & 7237       & 4.78E$-$02 & 3.73E$-$03 & 2.16E$-$02 \\
                He I          & 7281       & 1.60E$-$02 & 1.01E$-$02 & 3.80E$-$04 \\
                {[}O II{]}    & 7320       & 8.25E$-$01 & 5.00E$-$01 & 1.70E$+$00 \\
                {[}Ar III{]}  & 7751       & 1.03E$-$01 & 4.53E$-$02 & 3.70E$-$02 \\
                He II         & 8237       & 1.55E$-$02 & 9.90E$-$03 & 3.53E$-$04 \\
                H  I          & 8545       & 2.08E$-$02 & 6.25E$-$03 & 2.36E$-$03 \\
                H  I          & 8665       & 1.11E$-$02 & 9.42E$-$03 & 3.40E$-$04 \\
                H  I          & 8750       & 6.62E$-$03 & 1.18E$-$02 & 3.04E$-$04 \\
                H  I          & 8863       & 8.00E$-$03 & 1.52E$-$02 & 5.74E$-$04 \\
                {[}S III{]}   & 9069       & 1.46E$-$01 & 1.88E$-$01 & 2.84E$-$02 \\
                H I           & 9229       & 8.23E$-$02 & 2.50E$-$02 & 3.66E$-$02 \\
                He II         & 9345       & 2.85E$-$02 & 1.42E$-$02 & 2.25E$-$03 \\
                {[}S III{]}   & 9531       & 2.22E$-$01 & 4.72E$-$01 & 1.00E$+$00 \\

				\hline
		    \end{tabular}}
	\end{center}
	$^{a}$Relative to H$\beta$
\end{table}

\subsection{Photoionization Modeling of optical and NIR spectra}

We have used the photoionization code \textsc{cloudy}, C23.01 \citep{chatzikos2023} to model the early phase NIR spectrum from \citet{raj2011nirnova} and the nebular phase optical spectrum of V5579 Sgr taken on 2008 May 2 and 2011 June 2, 15 and 1139 days, respectively, after the outburst. By choosing the "pre-dust" (15 days) and "nebular phase" (1139 days) spectrum, we focused on a phase with minimal interference from dust, allowing us to better study the emissions from various elements. The photoionization code \textsc{cloudy} employs detailed microphysics to simulate the physical conditions of non-equilibrium gas clouds under the influence of an external radiation field. \textsc{cloudy} uses 625 species, including ions, molecules, and atoms, and employs five distinct databases for spectral line modeling: \textsc{stout}, \textsc{chianti}, \textsc{lamda}, H-like and He-like isoelectronic sequences, and the H{$_2$} molecule. All significant ionization processes, such as photoionization, Auger processes, collisional ionization, charge transfer, and various recombination processes (radiative, dielectronic, three-body recombination, and charge transfer), are included self-consistently. For a specified set of input parameters, \textsc{cloudy} predicts the intensities and column densities of a large number of spectral lines ($\sim 10^4$) across the electromagnetic spectrum, particularly for non-local thermodynamic equilibrium (NLTE) gas clouds, by solving the equations of thermal and statistical equilibrium in a self-consistent manner. This simulated spectrum can then be compared with the observed spectrum to estimate the physical and chemical characteristics such as the temperature and luminosity of the central ionizing source, the density and radii of the ejecta, and the chemical composition of the ejecta. For a detailed description of \textsc{cloudy}, see \citet{ferland2013}. 
\par
We consider a central ionizing source surrounded by a spherically symmetric ejecta whose dimensions are estimated by inner ($r_{in}$) and outer ($r_{out}$) radii. The central ionising source is assumed to have a blackbody shape with temperature T (K) and luminosity L (erg s$^{-1}$). We assume the surrounding ejecta to be clumpy and use the filling factor parameter to set the clumpiness in our \textsc{cloudy} models. The filling factor $f(r)$ varies with radius according to the following relation:
\begin{equation}
    f(r) = f(r_{in})(r/r_{in})^{\beta}
\end{equation}
where, $r_{in}$ is the inner radius, and $\beta$ is the power law exponent. The filling factor in novae ejecta typically ranges from 0.01 to 0.1 \citep{Shore_2008}. During the early stages, the value is often at the higher end, around 0.1, but decreases over time \citep{ederoclite2006}. In our analysis, we adopted a filling factor of 0.1 for the day 15 spectra, while for the nebular phase spectra, we treated the filling factor as a free parameter.  The ejecta density is determined by the total hydrogen number density parameter, $n(r)$ (cm$^{-3}$), which varies with the radius of the ejecta according to the following relation:
\begin{equation}
    n(r) = n(r_{in})(r/r_{in})^{\alpha}
\end{equation}
where, $n(r)$ is the density at radius $r$, and $\alpha$ is the exponent of power law. In the present study, we have used $\alpha = -3$ for a shell undergoing ballistic expansion and $\beta = 0$, consistent with the values used in previous studies of novae \citep[e.g.,][and references therein]{2010AJHelton,raj2018cloudy2676,2020MNRASPavana,2022ApJPandey,2022MNRASPandey,2024MNRASHabtie}. We set the chemical composition of the ejecta by the abundance parameter in \textsc{cloudy}. We varied the abundances of only those elements whose emission lines are present in the observed spectra, while we kept the remaining elements at their solar values \citep{2010Ap&SSGrevesse}.
\par
In our \textsc{cloudy} models, we used $T$, $L$, $n(r)$, and abundances as free input parameters. We employed a wide range of input parameters, varying them in smaller steps throughout the sample space, to generate a comprehensive set of synthetic spectra. We then match these model-generated spectra with the observed spectra to constrain the crucial characteristics of the nova system. After achieving a comparable match with our observations, we referred our attention to the emission lines that originate from elements such as carbon, oxygen, and nitrogen. Furthermore, we scrutinized the faint lines of sodium, argon, and iron by adjusting the abundances of all elements simultaneously, while allowing for slight variations in temperature, luminosity, and density parameters.

\begin{figure}
 \includegraphics[width=1\columnwidth]{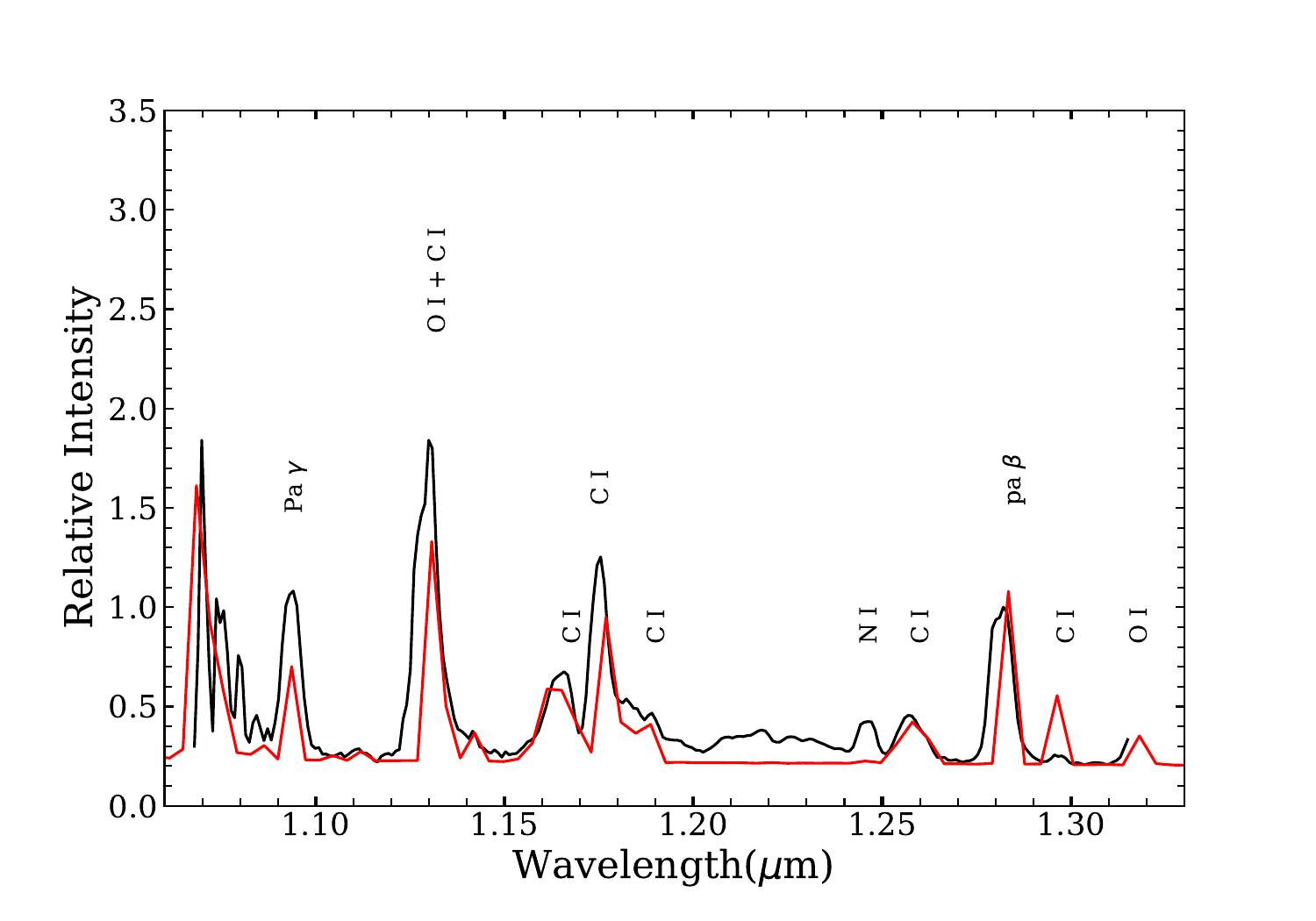}
 \includegraphics[width=1\columnwidth]{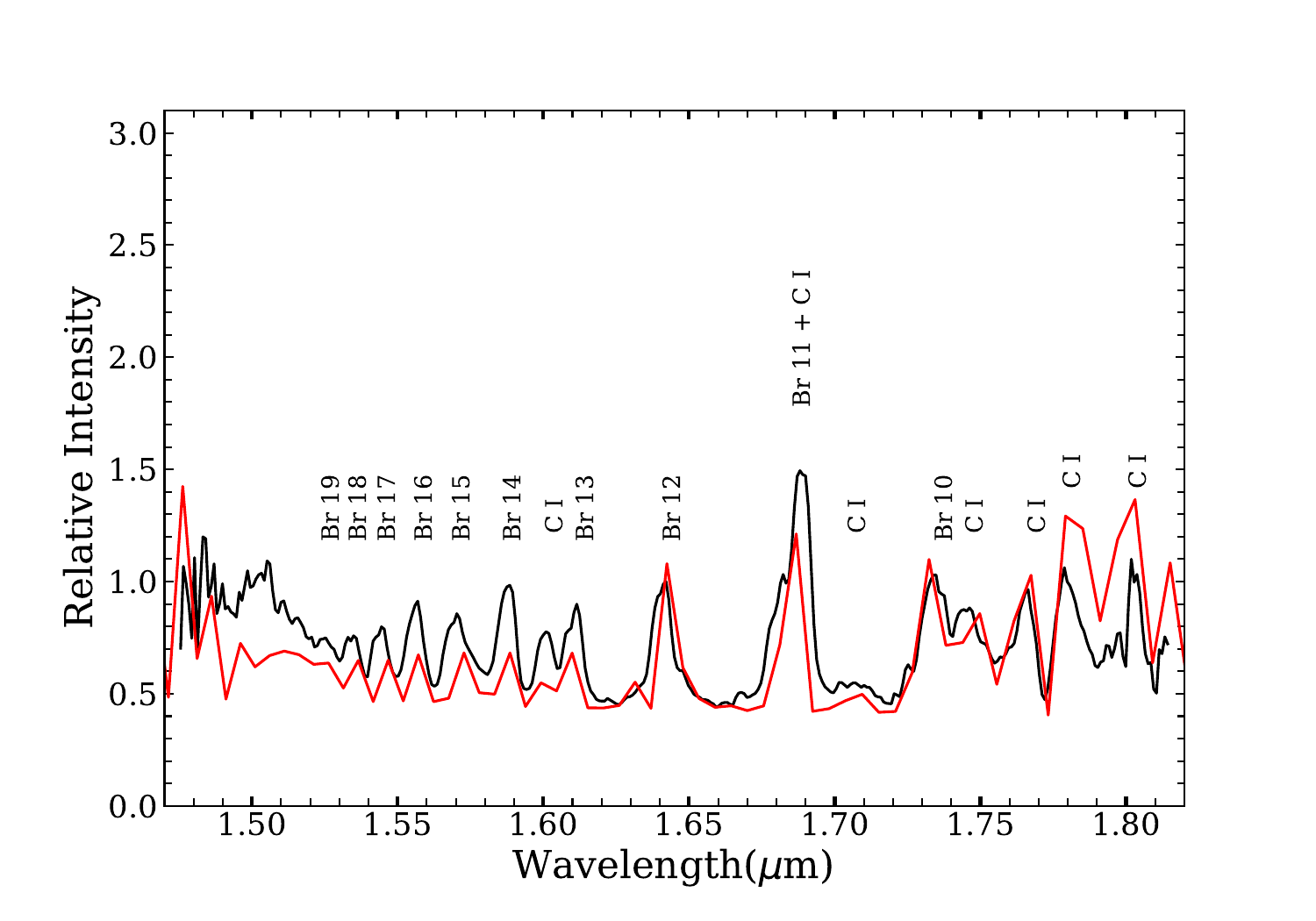}
 \includegraphics[width=1\columnwidth]{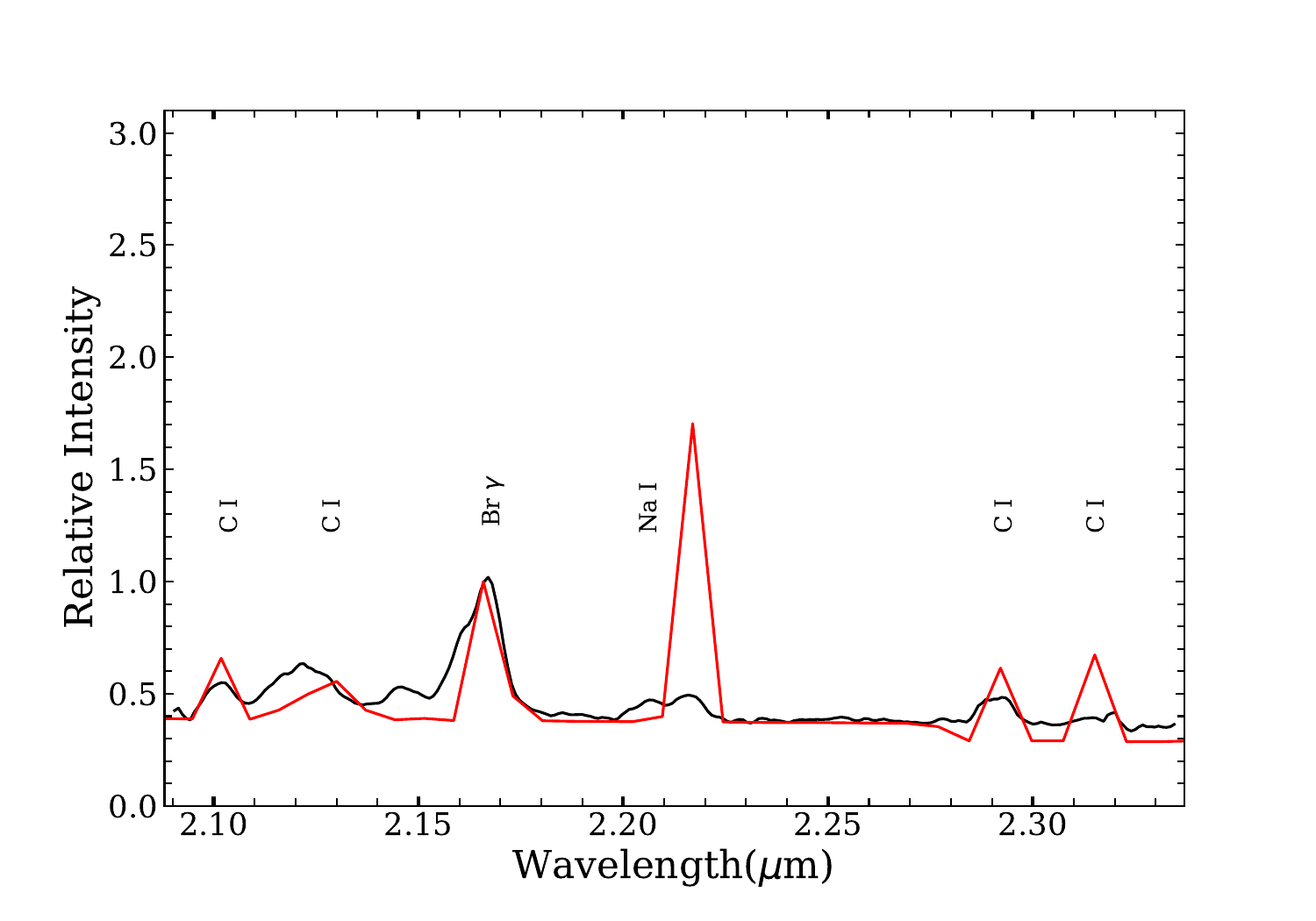}
\caption{Best-fit \textsc{cloudy} synthetic spectrum (red line) plotted over the observed spectrum (black line) for JHK bands obtained on 2008 May 3 (day 15). }
	\label{d13_jhk}
\end{figure}

\begin{figure*}
	\includegraphics[scale=0.45]{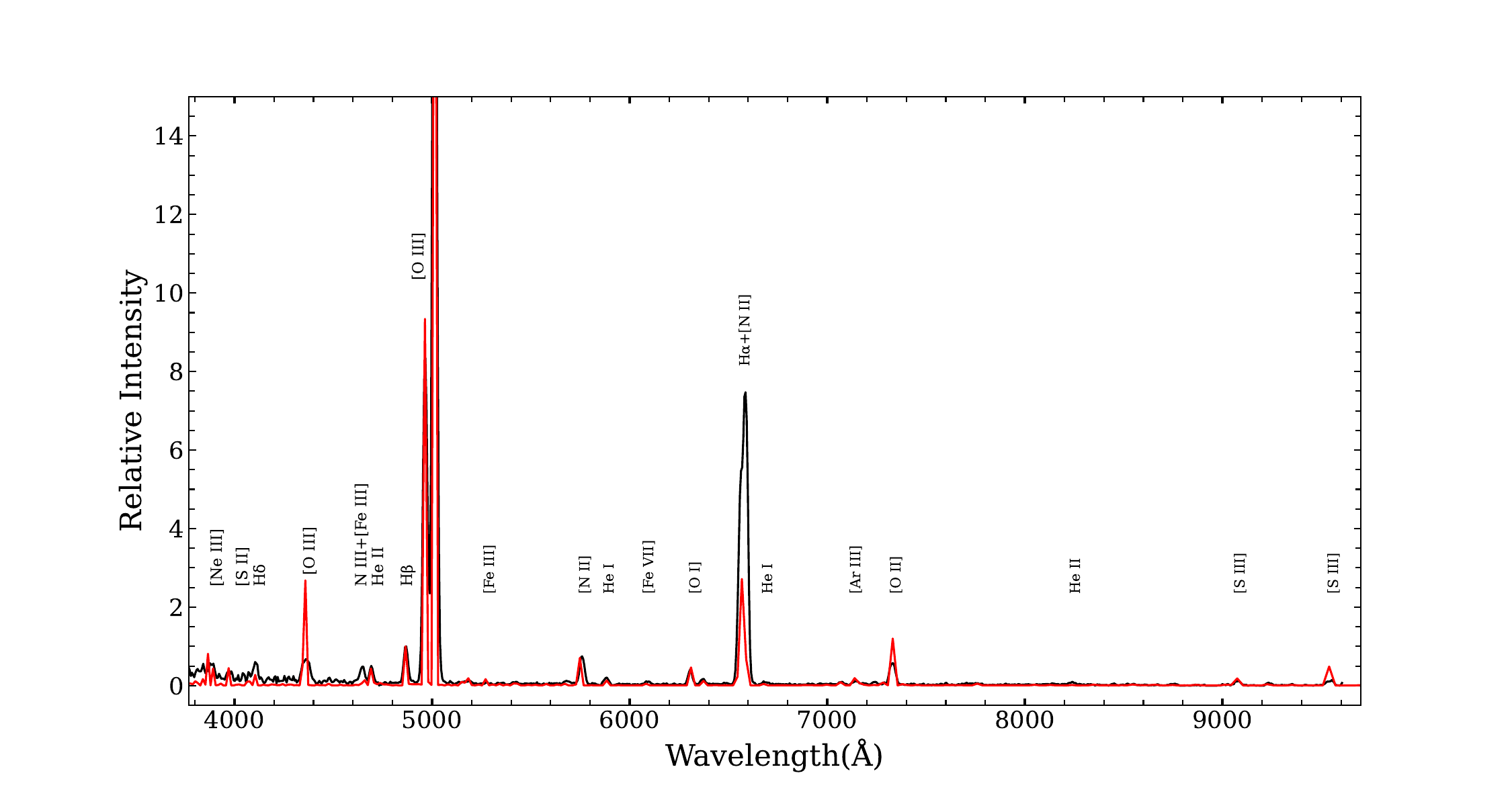}
	\caption{Best-fit \textsc{cloudy} synthetic spectrum (red line) plotted over the observed spectrum (black line) of V5579 Sgr obtained on 2011 June 2 (day 1139).}
	\label{d1139_1d}
\end{figure*}

\par
Nova ejecta are known to have an inhomogeneous density distribution, evidenced by the
presence of clumps and varying densities within the expanding material \citep{BodeEvansBook2008}.
In our initial modeling efforts, we found that models with a single uniform density were insufficient to generate emission lines across all transitions. To address this issue, we adopted a two-component density model. The component characterized by high ejecta density produced permitted emission lines in the spectra, while the component with lower ejecta density accounted for the forbidden emission lines. This low density allowed the ionizing photons to penetrate deeper into the shell, causing it to become more ionized and hotter \citep{shore2003early}.

It should be noted that our CLOUDY modeling assumes a simple spherically symmetric
geometry. In reality, the actual geometry of nova ejecta can be more complex, such as bipolar or
irregular shapes. CLOUDY, being a 1D photoionization code, cannot simulate bipolar and other
complex geometries. The assumption of spherical symmetry might oversimplify the actual structure of the
ejecta, and this complexity can affect the ionization and emission properties. We were unable to match the H$\alpha$ line as it was very strong and blended with the [N II] line. Hydrogen Balmer
emission lines generally form in the outer region of the ejecta, and our assumption of spherical symmetry might be
affecting the modeled strength of the H$\alpha$ emission line \citep{schwarz2001massejecta}.We may be able to test this if the outer regions of the ejecta deviate
from this assumption of spherical symmetric geometry. However, the full treatment of non-spherical
bipolar geometry is currently out of the scope of this paper. In our future work, we will
consider using 3D photoionization codes which can handle non-spherical geometries. So we will ignore this line, but other lines are generated and match very well.

We also set the covering factor, a measure of how much of the material we can see, for both of these regions in a way that when we add them up, it never exceeds unity.
To simplify, we keep most of the details constant for both regions, reducing the number of parameters we need to adjust. The ratios of the lines we model are calculated by adding up the ratios from each region, each multiplied by its covering factor. Finally, we set the inner and outer boundaries of the ejected material based on how fast it is expanding, which we calculate from the width of the emission lines, similar to a previous study by \citet{raj2018cloudy2676}.  
\\
As \textsc{cloudy} uses many parameters which are interdependent to generate a spectrum, it is difficult to validate the final spectrum visually. Thus, the best fit model is obtained by calculating $\chi^{2}$ and reduced $\chi^{2}$ :
\begin{align}
\begin{split}
\label{eqn2}
\chi^{2} = \sum_{i=1}^{n} \dfrac{(M_{i} - O_{i})^{2}} {\sigma^{2}_{i}}\\
\chi^{2}_{red} = \dfrac{\chi^{2}}{\nu}
\end{split}
\end{align}
\\
M$_{i}$ and O$_{i}$ represent the modeled and observed line ratios, respectively. The symbol $\sigma_{i}$ denotes the error in the observed flux ratio. The degrees of freedom, denoted by $\nu$, are calculated as n - n$_{p}$, where n is the number of observed lines, and n$_{p}$ is the number of free parameters.
The observed and modeled flux ratios with the $\chi^{2}$ values are given in Tables \ref{nir_table} and \ref{optical_table}. In general, the error ($\sigma$) typically falls within the range of 10\% to 30\%. This range depends on several factors, including the strength of the spectral line relative to the continuum, the possibility of blending with other spectral lines, uncertainties in de-reddening value, and errors in the measured line flux \citep{2010AJHelton,Woodward_2024}. We considered $\sigma$ value of 20\% to 30\% for the present study.

To make sure our predicted brightness matches the actual observed brightness after accounting for reddening, we assume that V5579 Sgr is about 5.6 kpc away. The measured line brightness is adjusted for reddening, specifically with a value of E(B - V) = 0.7 mag, as estimated in section \ref{ outburst luminosity, reddening and distance}. We then compare the adjusted measured brightness with our model's output to calculate  $\chi^{2}$, which helps us find the best fit.

To minimize errors related to how we measure brightness at different wavelengths, we use the ratios of modeled and observed prominent hydrogen lines, such as H$\beta$ in the optical region, P$\beta$ in J band, Br12 in H band and Br$\gamma$ in K band for calibration. This approach helps us to fine-tune our model and improve the accuracy of our predictions.

The abundance values and other parameters obtained from the model are listed in Tables \ref{nir_parameters} and \ref{optical_parameters}, presented on a logarithmic scale. The results suggest that the levels of helium, oxygen, carbon, and nitrogen are higher compared to solar values. These values are approximations because the calculations rely on only three or four observed lines. The accuracy of abundance solutions is influenced by changes in the opacity and conditions of the ejected material. Despite significant uncertainties, a model with a low chi-square provides a reasonable estimate of abundances. 
However, the model provides estimates for various parameters, including temperature, luminosity, density, and opacity (as detailed in Tables \ref{nir_parameters} and \ref{optical_parameters}). These estimates fall within the expected range of values for Fe-II novae, even though they should be considered as rough approximations due to the limitations mentioned.
The best-fit modeled spectrum (red line) with the corresponding dereddened observed spectrum (black line) is shown in Figs. \ref{d13_jhk} and \ref{d1139_1d} in JHK and optical region, respectively.
We have calculated the mass of the ejecta within the model shell following \citet{schwarz2001massejecta}:
\begin{equation}
M_{\text{shell}} = n (r_\text{in}) f(r_\text{in})\int_{R_{\text{in}}}^{R_{\text{out}}}\left(\frac{r}{r_\text{in}}\right)^{\alpha+\beta} 4\pi r^2 dr
\end{equation}
The values for density, filling factor, $\alpha$, and $\beta$ are adopted from the best-fit model parameters. Estimation of the total ejected shell mass entailed the multiplication of the mass in both density components (clump and diffuse) by their respective covering factors, followed by the addition of these products. The mass of the ejected hydrogen shell is estimated to be 1.67 $\times$ 10$^{-4}$ M$_{\odot}$ and 6.36 $\times$ 10$^{-4}$ M$_{\odot}$ using the modeling results of day 15 and 1139, respectively. These values are consistent with the theoretical values estimated for CNe \citep{gehrz1998nucleosynthesis}.

\subsection{Dust temperature and mass}\label{dust temperatre and mass}

A change in slope in the optical light curve about 20 days after the outburst clearly indicates the onset of dust formation \citep{raj2011nirnova}. We anticipate an increase in the fluxes of the NIR band during dust formation. The fact that the H-band flux shows a slight increase, while the K-band flux significantly rises around this time suggests that the dust is self-absorbed (optically thin) in the H-band. We observe a decline, rather than a rise, in the J-band flux during this time, as dust contributes primarily to the K band and beyond.
In the present study, we used day 15 and 1139 spectra for the photoionization modeling. Based on our best fit \textsc{cloudy} models, the carbon abundance (C/H) was estimated to be 15.5 relative to the solar value on day 15 and decreased to its solar value on day 1139. In contrast, the oxygen abundance (O/H) was estimated to be 32.2 on day 15 and decreased to 5.8 relative to the solar value during the subsequent phase, indicating that the ejecta could be rich in oxygen. The decrease in carbon abundance can be attributed to its depletion in dust grain formation. Theoretical hydrodynamic models of the TNR also suggest that the gas phase of the nova ejecta has a relatively higher abundance of oxygen (O). This is because an environment where oxygen (O) is higher than carbon (C) may also make it easier for carbon-rich dust grains to form \citep{starrfield2016thermonuclear}.

To estimate the outburst luminosity, we have used the ($\lambda F_{\lambda})_{max}$ values for optical and IR from \citet{raj2011nirnova}. Using the relation given by \citet{Gehrz_2008} that the outburst luminosity is equal to 4.11 $\times$ 10$^{17}$d$^{2}$ ($\lambda F_{\lambda})_{max}$ L$_{\odot}$. We calculate the outburst luminosity (L$_{O}$) on day 5 and infrared dust luminosity on day 26 (L$_{IR}$) to be 4.59$\times$10$^{5}L_{\odot}$ and 3.12$\times$10$^{4}$L$_{\odot}$, respectively, using distance d = 5.6 kpc. 
The inferred visible optical depth ($\tau = \frac{L_{\text{IR}}}{L_{\text{$O$}}}$) is calculated to be 0.07, indicating the presence of thin optical dust emission during the early dust formation phase. 
Similar values were previously determined for V1668 Cyg \citep{gehrz1988infrared} and V1831 Aql \citep{banerjee2018near}. 
\citet{banerjee2018near} suggested that small $\tau$ values imply either clumpy dust distribution, hence does not cover all angles of the sky towards the nova as seen by the observer or a homogeneous dust shell with insufficient material to achieve optical thickness.

\begin{figure*}[hbt!]
\begin{center}
\includegraphics[trim=0.3cm 0.5cm 0.5cm 0.25cm, clip, width=0.98\textwidth]{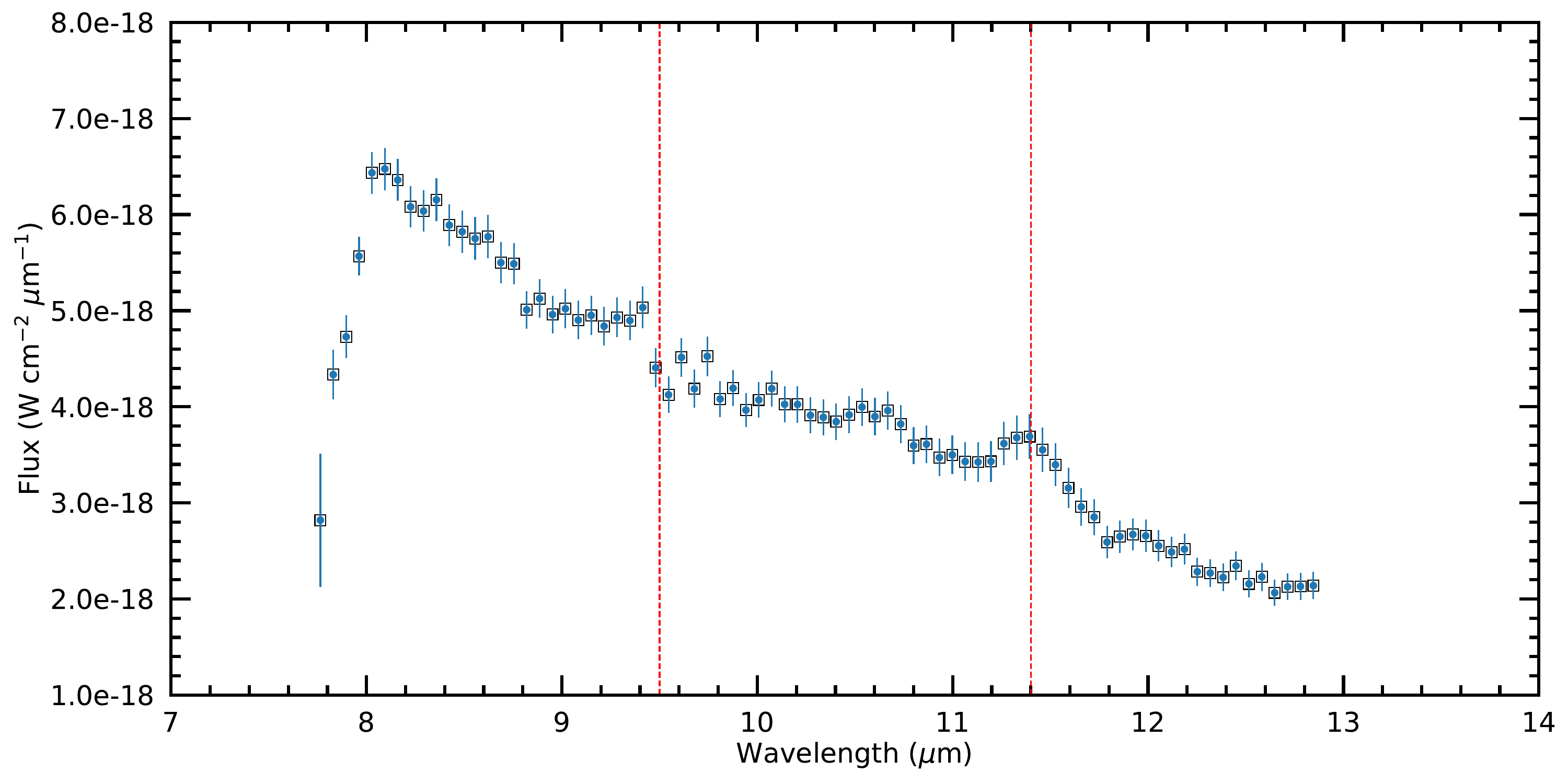}
\caption{The 7.70 to 14.00~$\mu$m low resolution $(R \simeq 120)$ spectrum of V5579 Sgr 
obtained on Gemini-S (+TReCS) on 2009 Sept 23. (Day +522.26). The vertical dashed red lines are the
\citep[PAHs;][]{allamandola1989interstellar} complexes detected in the NASA Spitzer spectra of V2362~Cygni. 
The slope of the observed TReCS spectrum of V5579~Sgr follow that of the Rayleigh-Jeans tail of a hot blackbody arising from condensed dust in the ejecta \citep[e.g.,][]{rudy2008v5579, raj2011nirnova}. }
\label{fig:trecs-spec-v5579sgr}
\end{center}
\end{figure*}

Dust condensed in the ejecta of V5579 Sgr as early as $\simeq +20~$d after outburst \citep{2008IAUC.8948....1R}.
The TReCs spectrum on $\simeq 522$~d (Figure~\ref{fig:trecs-spec-v5579sgr}) show that emission from
dust is still present at this late epoch. The Rayleigh Jean tail of the hot blackbody emission arising
from carbonaceous grains \citep[which have featureless spectral energy distributions in the infrared][]{draine2007infrared}
has spectral features superposed on the otherwise smooth continuum. In V5579~Sgr, there are spectral features evident 
in the spectral energy distribution at 9.5 and 11.4 $\mu$m (see Fig. \ref{fig:trecs-spec-v5579sgr}) associated with Polycyclic Aromatic Hydrocarbon (PAH) like features seen in the 
V2362~Cygni \citep{helton2011atypical} and in V705~Cas \citep{1997Ap&SS.251..303E}. V5579~Sgr thus joins
the growing list of novae where such materials have formed.

\begin{figure}
	\includegraphics[scale=0.35]{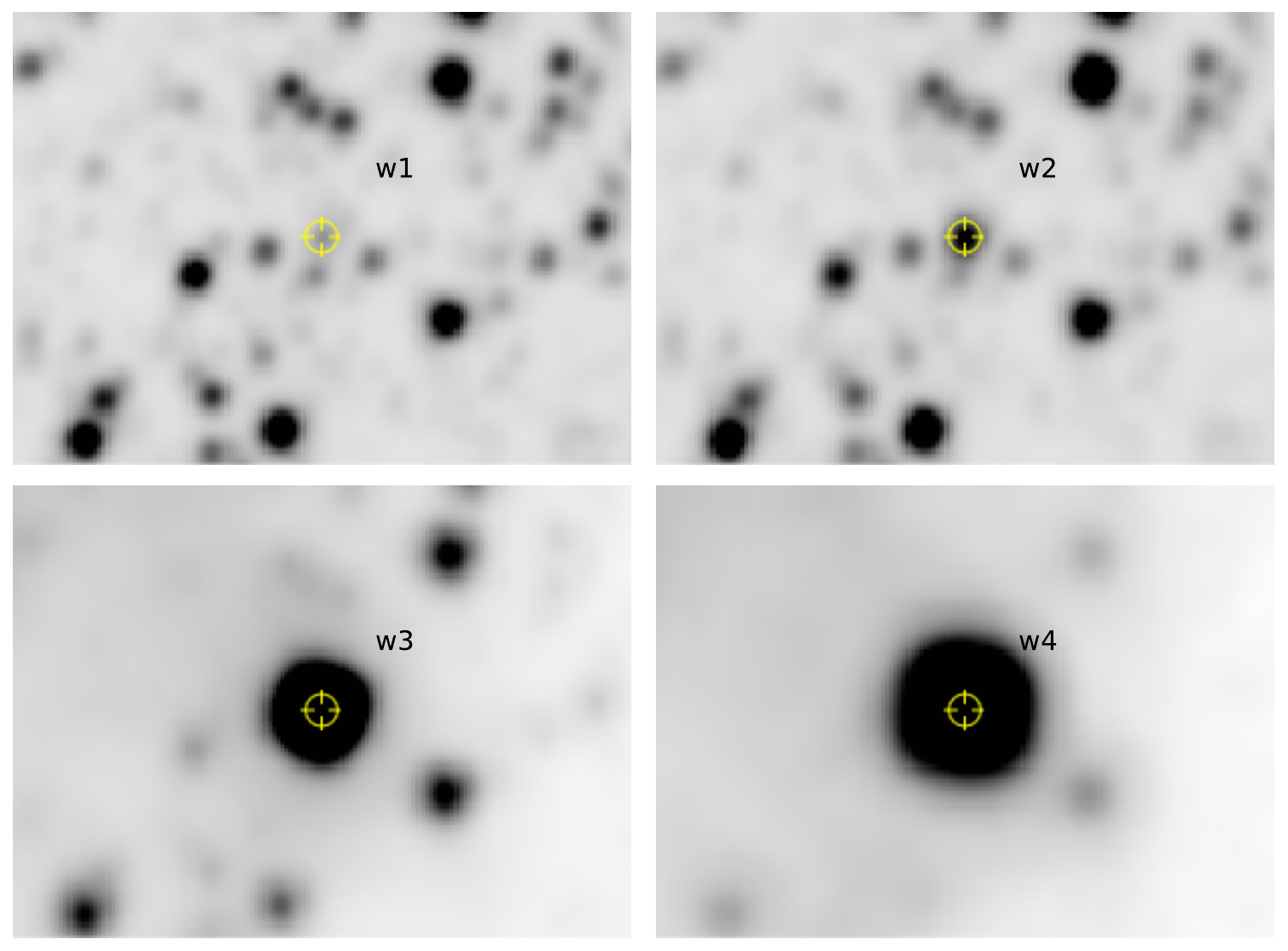}
	\caption{A mosaic of a 3x3 arc minute square field around V5579 Sgr. The source is detected in all 4 WISE bands: W1 (3.4~$\mu$m), W2 (4.6~$\mu$m), W3 (12~$\mu$m) and W4 (22~$\mu$m); the emission at the longer W3 and W4 bands is very
pronounced. The WISE images were taken in March 2010 (more details in section \ref{dust temperatre and mass}),
nearly 2 year since discovery . Although the nova had faded below
15 magnitudes in the V band by this time (see Fig. \ref{lc_optical}), it remained strikingly bright in the near
and mid IR due to emission from newly formed dust in the ejecta.}
	\label{wise_image}
\end{figure}

\begin{figure}
	\includegraphics[scale=0.35]{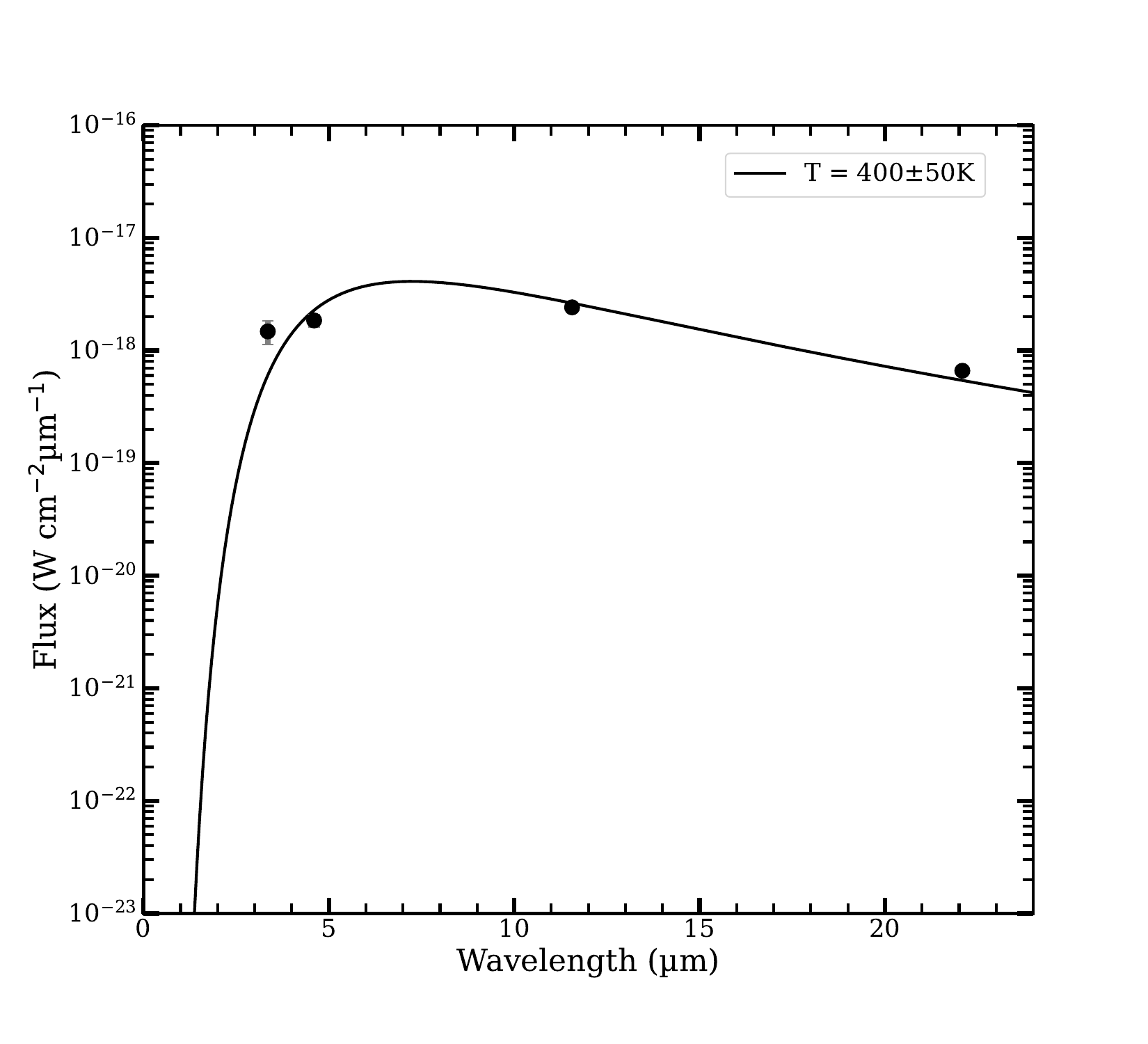}
	\caption{The SED shows a blackbody fit to the WISE data taken on 2010 March 22 with a temperature of about 400 K.}
	\label{SED_fit}
\end{figure} 

The observations from the Wide field Infrared Survey Explorer (WISE: \cite{wright2010wide}) also support emission from the dust at longer wavelengths (see Fig. \ref{wise_image}).
We estimate the temperature of the dust shell around 702 days after the outburst by using the WISE magnitudes as
400 $\pm$ 50 K. However, the temperature estimate for the dust shell
may have a large uncertainty, as we assume that the isothermal dust and we have used only four wavelengths to fit the SED where the dust could contribute at larger wavelengths beyond 22 $\mu$m. 

The mass of the dust shell is calculated from the SED on 22 March 2010 shown in Fig. \ref{SED_fit}. We estimate the dust mass following \citet{evans2017rise} and
\citet{banerjee2018near}, assuming that the grains are spherical and
that the dust is composed of carbonaceous material. 
Using the relations given by \citet{evans2017rise}, we find that the dust masses for amorphous carbon (AC) and graphitic carbon (GR) grains are as follows.
For optically thin amorphous carbon,
\begin{equation}
\frac{M_{dust-AC}}{M_{\odot}} \simeq 5.83 \times 10^{17} \frac{(\lambda f_{\lambda})_{max}}{T^{4.754}_{dust}}
\end{equation}
and for optically thin graphitic carbon,
\begin{equation}
\frac{M_{dust-GR}}{M_{\odot}} \simeq 5.19 \times 10^{19} \frac{(\lambda f_{\lambda})_{max}}{T^{5.315}_{dust}}
\end{equation} 
where we assume the distance $D$ = 5.6 kpc, the density of the carbon grains $\rho$ = 2.25 gm cm$^{-3}$, and ($\lambda$ $f_{\lambda}$)max is in unit of W m$^{-2}$. The dust mass, which is independent of grain size
\citep{evans2017rise}, is estimated to be $\sim$ 8.3$\times 10^{-8}M_{\odot}$  and
2.6$\times 10^{-7}M_{\odot}$ , for AC and GR grains, respectively.
AC grains provide a much better fit across a wider temperature range (400-1700 K) compared to graphitic carbon, which shows a reasonably good fit only within the narrower range of 700-1500 K \citep{blanco1983planck}. This finding suggests that the dust grains in the nova are probably composed mainly of AC. 
The value of masses show that a large amount of optically thick dust was formed after May 2008 in the nova ejecta. 

The dust mass might not be totally accurate for some reasons as the dust might have only formed in some parts of the ejecta, not everywhere, 
and the temperatures we measured might not be totally accurate. The dust's ability to emit light 
depends on composition and the size of the dust grain, and this can make a difference in how we see its temperature \citep{2003pid..book.....K}.

\section{Discussion}\label{discussion}
The observed NIR excess suggests that a large amount of optically thick dust was formed in the nova ejecta. After a peak was reached, the J-H and H-K colors started to show a downward trend, indicating either the destruction of the dust molecules due to radiation from the central source or the thinning of the dust shell as it expands and becomes less evident. A similar behaviour was seen in V2676 Oph where molecule formation before dust condensation was reported and had similar colors, J-K = 8, J-H = 4.5 and H-K = 3.8 mag \citep{raj2017v2676}. Dust formation in CNe is an intriguing phenomenon in astrophysics because the nova forms dust within several days to several tens of days after the explosion (TNR) and ejects the molecules and dust grains, including PAH, into galactic space \citep{gehrz1998nucleosynthesis}. Understanding the location of the dust condensation layer is very important to understand dust formation in CNe. 

Usually, the chemical composition of the stellar atmosphere determines the type of dust that forms in the stellar ejecta. Studies have shown that a higher ratio of carbon to oxygen (C > O) is necessary for the formation of silicon carbide (SiC) and amorphous carbon grains. This is because CO is the most stable molecule at T $<$ 2000K, at which grains usually form. As the CO molecule once formed in the ejecta, all of the remaining oxygen was trapped in the CO molecule, leaving only residual carbon available for grain formation. In contrast, when oxygen predominates over carbon (O > C), the carbon is locked up in CO formation, while remaining oxygen tends to combine with other elements to form oxides and silicates \citep{2004ASPCWaters}. However, it has been observed that some novae are capable of producing significant amounts of both carbon-rich and oxygen-rich dust grains. Examples of such cases are V705 Cas \citep{evans2005infrared}, V1065 Cen \citep{2010AJHelton}, V5668 Sgr \citep{gehrz2018temporal}, V1280 Sco \citep{2022ApJPandey}, etc. This may imply that the CO formation process does not reach saturation in the nova ejecta, thereby permitting the incorporation of both carbon and oxygen into the dust. On the other hand, this could also mean that there are differences in the amounts of carbon and oxygen in the ejecta, which would cause the carbon-to-oxygen (C:O) ratio to vary depending on where it is found. The exact mechanism for the presence of this bimodal dust in nova ejecta is not well known \citep{starrfield2016thermonuclear,sakon2016concurrent}.

V5579 Sgr belongs to the fast novae speed class (t$_2$ $\sim$ 9 days) with small fluctuations in the NIR light curve in the early phase followed by a decline in the late phase. In the optical light curve, from day 20 a decline is seen due to dust formation. 
Only nova Cyg 2006 (V2362 Cyg) and nova Aql 1995 (V1425 Aql) formed an optically thin dust with a similar value for t$_2$. But V2362 Cyg showed a pre-maximum rise lasting for about 2.8 days and also a second maximum at $\Delta$t = 239 d, after the optical maximum and
dust formed around $\Delta$t = 251 d. In the case of V1425 Aql there was an indication of dust in the ejecta around $\Delta$t = 22 d, while the present observations show that dust formation occurred even earlier around $\Delta$t = 15 d after the optical maximum in V5579 Sgr. Thus the IR data presented here for V5579 Sgr along with the results of \citet{1997AJkamath} on V1425 Aql indicate that the earliest
dust formation time scale for an Fe II nova lies in the range of 15 - 22 days.

Our \textsc{cloudy} modelling supports the notion of a nitrogen enrichment in the V5579 Sgr nova ejecta, likely due to proton capture during thermonuclear runaway (TNR). This finding aligns with observations of other novae, such as V705 Cas, where \citet{hauschildt1994early} reported an enhancement of heavy elements (particularly carbon, nitrogen, and oxygen) compared to solar abundances. \citet{shore2018spectroscopic} explored the potential consequences of a significantly elevated nitrogen-to-carbon (N/C) ratio, particularly its influence on dust formation in CO-type novae. The overall increase in heavy element abundance within the ejecta can potentially disrupt the thermal balance of the gas, fostering conditions that favor the formation of dust nucleation sites. As suggested \citet{ferland1978heavy}, a higher metallicity allows for more efficient cooling of the ejecta, facilitating the achievement of thermal equilibrium at lower kinetic temperatures.

\section{Summary}\label{summary}
We have presented a comprehensive optical spectrophotometric and NIR photometric evolution. 
The important results of the analyses are summarized here.
\begin{enumerate}
 \item Using optical and NIR data from AAVSO and SMARTS, t$_2$ was estimated to be 9 $\pm$ 0.2 d, indicating that V5579 Sgr belongs to the class of fast novae.
\item The reddening E(B - V) was estimated to be about 0.7 $\pm 0.14$ and the distance to the nova was found to be 5.6 $\pm 0.2$ kpc. Using MMRD relation we have estimated the absolute magnitude of the nova M$_v$
 to be 9.3 $\pm 0.1$. 
 \item The mass of WD was estimated to be 1.16 M$_{\odot}$.
\item The spectral evolution clearly indicates that it belongs to the Fe II class of novae. The nova evolves in PfeAo sequence as per the classification given by \citet{williams1991evolution}. Coronal lines are absent in the optical spectrum during the late phase. 
\item We utilized the \textsc{cloudy} photoionization code to simulate the optical and near-infrared spectra of the dust-forming nova V5579 Sgr. From the best-fit model, we estimate various physical and chemical parameters of the system. Our abundance analysis shows that the ejecta are significantly enhanced relative to solar, O/H = 32.2, C/H = 15.5 and N/H = 40.0 in the early decline phase and O/H = 5.8, He/H = 1.5 and N/H = 22.0 in the nebular phase.
\item The low-resolution mid-IR Gemini spectra obtained 522 days since discovery showed PAH-like features. Using the WISE magnitudes, we found that the temperature of the dust is approximately 400$\pm$ 50 K on day 702.
The dust mass is estimated to be $\sim$ 8.3$\times 10^{-8}M_{\odot}$  and
2.6$\times 10^{-7}M_{\odot}$, for AC and GR grains, respectively. 

\end{enumerate}

\begin{acknowledgement}
The authors would like to thank the referee for critically reading the manuscript and providing valuable suggestions to improve it. We acknowledge with thanks the variable star observations from the AAVSO International Database contributed by observers worldwide and used in this research. We also acknowledge the use of SMARTS data. R Pandey acknowledges the Physical Research Laboratory, Ahmedabad, India, for her post-doctoral fellowship. 

Based on observations obtained at the international Gemini Observatory, a program of NSF NOIRLab, acquired through the the Gemini Science Archive prior to December 2015 [GS-2009B-Q68], which is managed by the Association of Universities for Research in Astronomy (AURA) under a cooperative agreement with the U.S. National Science Foundation on behalf of the Gemini Observatory partnership: the U.S. National Science Foundation (United States), National Research Council (Canada), Agencia Nacional de Investigaci\'{o}n y Desarrollo (Chile), Ministerio de Ciencia, Tecnolog\'{i}a e Innovaci\'{o}n (Argentina), Minist\'{e}rio da Ci\^{e}ncia, Tecnologia, Inova\c{c}\~{o}es e Comunica\c{c}\~{o}es (Brazil), and Korea Astronomy and Space Science Institute (Republic of Korea).\end{acknowledgement}




\printendnotes

\printbibliography

\end{document}